\documentclass{aa}

\usepackage{graphicx}
\usepackage{amsmath}
\usepackage{amssymb}
\usepackage{natbib}
\bibpunct{(}{)}{;}{a}{}{,}

\begin{document}

\title{Hamiltonian variational reconstruction of the 3D magnetic geometry of relativistic jets}

\titlerunning{H-MOG: Hamiltonian reconstruction of jet magnetic geometry}
\authorrunning{F. Buffoli}

\author{F.~Buffoli\inst{1}}

\institute{Universit\`a degli Studi di Brescia, Via San Faustino 54, 25123 Brescia, Italy\\
\email{fabio.buffoli@unibs.it}}

\date{}

\abstract
{Every resolved image of a relativistic jet encodes its magnetic field, yet no
image records which way the field points. Synchrotron intensity and linear
polarization probe the field only through even combinations of its
components.}
{We introduce H-MOG (Hamiltonian Magnetic Optimization of Geometry), a
variational method that reconstructs the 3D field from two projected
observables, the jet width $W(z)$ and the linear polarization $p(z)$, and we
determine exactly which properties of the field these observables fix.}
{We validate H-MOG on synthetic jets and apply it to ten general relativistic
magnetohydrodynamic (GRMHD) simulations of magnetically arrested and standard
accretion flows at five black-hole spins. We test four routes to recover the
field sense: a Faraday rotation-measure term, full 3D sampling, a spin-based
prior, and resolved multi-frequency polarimetry with internal Faraday
rotation, including ray-traced GRMHD images.}
{For a jet viewed side-on, $W$ and $p$ are exactly invariant under field
reversal, $\mathbf{B}\to-\mathbf{B}$, and under reflection through the plane
of the sky, which reverses the handedness of the helix. Within this
four-fold degeneracy the geometry is recovered, with
$\langle\cos\rangle\simeq0.94$--$1.00$ on synthetic jets and $0.76$--$0.98$
on the GRMHD library up to the mirror image; the image selected is inherited
from the initial guess. In the simulations the poloidal field has the same
direction in both jets at every spin, while the handedness of the helix
follows the sign of the spin. Side-on, the rotation measure fixes the sign
of the toroidal field but not that of the poloidal field, at any grid
resolution, and the spin prior fails because spin does not set the poloidal
sign. Internal Faraday rotation breaks all four images, but in ray-traced
GRMHD images it separates them only weakly.}
{From width and linear polarization alone, field sense and helical handedness
are undetermined. For a side-on jet, the rotation measure fixes the toroidal
sign; the poloidal sense also requires the rotation sense of the field lines,
a viewing angle away from $90^\circ$, or circular polarization. With the rotation measure and the rotation sense, a
multi-start minimization recovers the poloidal sense of a GRMHD jet.}


\maketitle


\section{Introduction}
\label{sec:intro}

Relativistic jets from active galactic nuclei are launched by magnetic
fields. The standard picture has two variants. In one, the field threads
the horizon of a spinning black hole and extracts rotational energy through
the Poynting flux \citep{bz77}. In the other, it is anchored in the
accretion disc and drives a magnetocentrifugal wind \citep{blandford_payne82}.
Both mechanisms require an ordered poloidal field, and the efficiency of
the first scales steeply with the magnetic flux threading the hole and
with the black-hole spin \citep{tchekhovskoy11,mckinney12}. The flux
itself is regulated by the accretion state. In the magnetically arrested
disc (MAD) state the horizon saturates with poloidal flux and the jet is
powerful and well collimated \citep{narayan2003,tchekhovskoy11}, while in
the standard and normal evolution (SANE) state the field is weaker and
more turbulent \citep{narayan2012,sadowski2014}. General relativistic
magnetohydrodynamic (GRMHD) simulations have made these pictures
quantitative, reproducing the launching, collimation and acceleration of
jets across a range of spins and magnetizations
\citep{gammie2003,mckinney2006,porth19}, and they are now the standard
tool for interpreting horizon-scale data \citep{eht2021_viii,dhruv2025}.
The extraction mechanism itself has been shown to operate beyond the Kerr
spacetime, in generic stationary and axisymmetric black-hole spacetimes
\citep{chatterjee2025,camilloni2026}.

M87 is where these ideas meet observation most directly. Its proximity and
large black-hole mass yield the highest angular resolution in
gravitational-radius units of any jet \citep{eht19}. Decades of very long
baseline interferometry (VLBI) have
now mapped its structure from kiloparsec scales down to a few gravitational
radii \citep{asada2011,hada2011,doeleman2012,akiyama2015,kim2018,walker2018}.
The jet collimates from a parabolic to a conical profile near the Bondi
radius \citep{asada2011,nakamura2018}, a shape now traced to within the
sphere of influence and compared directly with the outermost streamlines
of force-free jet models \citep{nakamura2018}. The Event Horizon
Telescope (EHT) has resolved the ring of emission around the black hole
\citep{eht19} and, in polarized light, has constrained the magnetic
field order at the jet base \citep{eht2021_vii,eht2021_viii}, favouring
a magnetically arrested state. Linear polarization has since been
measured for Sgr~A* as well \citep{eht2024sgra}, and circular
polarization --- a rarer and harder measurement --- has now been detected
near the M87 horizon \citep{eht2023m87circ}. The kinematics of the M87
jet, from sub-luminal inner regions to superluminal knots downstream,
fill in the dynamical picture \citep{mertens2016,walker2018}.

None of these observations, however, delivers the 3D magnetic field directly.
The geometry that shapes the jet is encoded there: how the poloidal and
toroidal components wind together, how the flow collimates, where it is
prone to instability \citep{mckinney12,tchekhovskoy11,mizuno2014}. But
this geometry is not read off straightforwardly from the image. What reaches the
observer is synchrotron radiation, whose total intensity scales with the
magnetic energy density, and linear polarization, which traces the
projected field orientation but not its sense; both are integrated along
the line of sight. The field that produced an image is therefore not
uniquely fixed by it. Recovering a 3D field from projected data is an
inverse problem, and like most inverse problems in astrophysics it is
ill-posed: many fields reproduce the same observables. The common
responses are to fit a low-dimensional parametric model, which builds the
answer into its assumed symmetry, or to map the Faraday rotation measure
(RM) and infer a helical component from a transverse gradient
\citep{zavala_taylor2004,broderick_mckinney2010,homan2009}. Neither
reconstructs the field as a free function on a 3D grid, and neither states
precisely what about the field is determined by the data and what is not.

Part of the problem goes beyond being ill-posed: it is strictly
non-identifiable. When observables depend on the field only through even
functions --- energy density $\propto|\mathbf{B}|^2$, polarization fraction
--- then $\mathbf{B}$ and $-\mathbf{B}$ give identical data. When, in
addition, they are integrals along the line of sight, a configuration and
its mirror image through the plane of the sky give identical data too. No
procedure that uses only those observables can prefer one member of such a
pair over the other. These are properties of the observables, not of any
particular algorithm. What they hide is physically loaded: the direction of
the poloidal field and the handedness of its toroidal winding, which encodes
the sense of rotation transmitted from the hole or disc
\citep{bz77,tchekhovskoy11}.

The distinction we draw is between what the data determine and what they
cannot --- not between easy and hard. We will show that from the width
profile and polarization fraction one recovers the unsigned 3D geometry
of the field --- collimation, degree of winding, opening of the field lines --- without
a parametric model, and we prove that the sense of the poloidal field and the
handedness of the helix are not determined by these observables.

This paper has three aims. We first introduce H-MOG (Hamiltonian Magnetic
Optimization of Geometry): a variational method
that reconstructs 3D jet magnetic geometry as a free field on a lattice,
regularized by a physically motivated Hamiltonian prior and optimized
with automatic differentiation. Second, we measure the accuracy of this
reconstruction on synthetic jets and on the public GRMHD library of
\citet{dhruv2025} --- ten simulations spanning both accretion states and
five black-hole spins. Third, we state the discrete degeneracies of the
problem formally, test whether a rotation-measure term, full 3D sampling, or
a spin-based prior can lift them, and identify in each case what limits
them.

Section~\ref{sec:method} describes the variational framework, the
Hamiltonian prior, the forward model, the numerical setup, and the formal
statement of the degeneracies. Section~\ref{sec:data} describes the
GRMHD library and how the field cubes and observables are extracted.
Section~\ref{sec:synth} validates the method on synthetic jets with known
truth. Section~\ref{sec:geom} presents the geometric reconstruction across
the ten GRMHD models. Section~\ref{sec:sign} tests the routes to recover the
field sense. Section~\ref{sec:discussion} discusses the nature of the limit,
its dependence on the viewing angle and the observational routes that could
lift it, and Section~\ref{sec:conclusions} concludes.

\section{Method: Hamiltonian variational inversion}
\label{sec:method}

\subsection{Variational framework}
\label{sec:vframe}

The magnetic field lives on a regular Cartesian lattice. Unless stated
otherwise the lattice has $N^3$ cells with $N=96$ and half-width
$L=30\,r_{\rm g}$ around the black hole, where $r_{\rm g}=GM/c^2$; in
Sect.~\ref{sec:sign} we also use a finer, anisotropic lattice around the jet
funnel. Rather than storing three
Cartesian components directly, we split the field into a non-negative
amplitude $A(\mathbf{x})\ge 0$ and a unit direction field
$\mathbf{S}(\mathbf{x})$ ($|\mathbf{S}|=1$):
\begin{equation}
\mathbf{B}(\mathbf{x}) = A(\mathbf{x})\,\mathbf{S}(\mathbf{x}).
\label{eq:split}
\end{equation}
This split is natural for the problem: the amplitude carries the emission
and hence the jet width, while the direction carries the pitch and hence
the polarization. The two are reconstructed jointly.

The reconstruction is posed as the minimization of a cost functional
\begin{equation}
\mathcal{L}[A,\mathbf{S}] = \mathcal{L}_{\rm data}[A,\mathbf{S}]
+ \alpha_{\rm prior}\,\mathcal{H}[\mathbf{S}]/N_{\rm cells}
+ w_{\rm div}\,\mathcal{L}_{\rm div}[A,\mathbf{S}],
\label{eq:loss}
\end{equation}
where $\mathcal{L}_{\rm data}$ measures the mismatch between the
observables computed from the candidate field and the target observables,
$\mathcal{H}$ is the Hamiltonian prior of Sect.~\ref{sec:prior},
$\mathcal{L}_{\rm div}$ penalizes the divergence of the field, and
$\alpha_{\rm prior}$ and $w_{\rm div}$ set the relative weights. We use a
variational formulation, not a trained network,
because we want a solution that satisfies the physics for this specific
object, with no dependence on a training set and no risk of importing
biases from one.

\subsection{Hamiltonian prior}
\label{sec:prior}

The data term alone does not determine the field uniquely; a prior is
required to select a physically sensible solution among those compatible
with the observables. We use a prior $\mathcal{H}$ built from terms that
encode the physics a relaxed, collimated jet field is expected to
satisfy. It is the sum of four contributions,
\begin{equation}
\mathcal{H} = \mathcal{H}_{\rm frame}
+ \lambda\,\mathcal{H}_{\rm kink}
+ \alpha\,\mathcal{H}_{\rm ext}
+ \beta\,\mathcal{H}_{\rm shear},
\label{eq:ham}
\end{equation}
each acting on the unit direction field $\mathbf{S}$.

The \emph{anti-kink} term $\mathcal{H}_{\rm kink}=\sum|\nabla\mathbf{S}|^2$
penalizes small-scale variation of the field direction. It encodes the
tendency of the field to organize on large scales, suppressing current-driven
kinking that would otherwise appear \citep{mizuno2014}. Next, the
\emph{external-pressure} term $\mathcal{H}_{\rm ext}=\sum(\theta-\theta_{\rm t}(z))^2$
acts on the local opening angle $\theta=\arccos|S_z|$. This pulls the field
towards a collimation profile $\theta_{\rm t}(z)$ expected for a jet confined
by an external medium \citep{asada2011,nakamura2018}. The \emph{shear-layer}
term $\mathcal{H}_{\rm shear}=\sum w(R)\,(|S_z|-\mu_{\rm t})^2$ acts in a
cylindrical shell of radius $R_0$; it encourages the field inclination to
approach a target value in that shell. Finally, a \emph{frame-dragging}
term $\mathcal{H}_{\rm frame}=-\sum h\,S_z$ can bias the axial field
towards a prescribed sense. It is switched off ($h=0$) in all
reconstructions except the spin-prior test of Sect.~\ref{sec:sign_spin},
so the field sense is never imposed by the prior.

We additionally enforce the solenoidal constraint through the penalty
$\mathcal{L}_{\rm div}=\langle(\nabla\!\cdot\!\mathbf{B})^2\rangle$
on the full field, so that the reconstructed field is as divergence-free as
the discretization allows \citep{toth2000}. The explicit lattice expressions
and weights for all terms are given in Appendix~\ref{app:prior}. With $h=0$
every term depends on $\mathbf{S}$ only through $|S_z|$, through squared
gradients or through $|\mathbf{B}|$; Sect.~\ref{sec:degeneracy} shows that
this makes the prior share the symmetries of the data.

\subsection{Forward model and observables}
\label{sec:forward}

From a candidate field we compute two projected observables that stand in
for what VLBI provides: the transverse jet width and the linear
polarization degree, both as functions of distance $z$ along the jet
axis. Throughout this paper the observer looks along the $y$ axis, so the jet
axis $z$ lies in the plane of the sky: all results refer to a fixed viewing
angle of $90^\circ$ (a jet seen side-on). We discuss other inclinations in
Sect.~\ref{sec:disc_incl}.

We take the synchrotron emissivity proportional to the magnetic energy
density, $\varepsilon \propto |\mathbf{B}|^2$, with no frequency dependence
and no relativistic beaming. Integrating along the line of sight gives a
projected image $I(x,z)=\int \varepsilon\,dy$. At each height $z$ the jet
width $W(z)$ is defined from the second moment of the transverse brightness
profile,
\begin{equation}
W(z) = 2\sqrt{2\ln 2}\;
\left[\frac{\int (x-\bar{x})^2\,I(x,z)\,dx}{\int I(x,z)\,dx}\right]^{1/2},
\label{eq:width}
\end{equation}
i.e.\ the FWHM-equivalent width of the emission about its intensity-weighted
centroid $\bar{x}(z)=\int x\,I(x,z)\,dx\big/\!\int I(x,z)\,dx$. The linear polarization
follows from the projected field angle: with
$\chi = \arctan(B_z/B_x)+\pi/2$ the local plane-of-sky orientation of the
electric vector, we form, at each height,
\begin{equation}
\begin{aligned}
I(z)&=\iint \varepsilon\,dx\,dy,\\
Q(z)&=\iint \varepsilon\cos 2\chi\,dx\,dy,\qquad
U(z)=\iint \varepsilon\sin 2\chi\,dx\,dy,
\end{aligned}
\label{eq:stokes}
\end{equation}
and define the polarization degree of the slice
\begin{equation}
p(z) = \frac{\sqrt{Q(z)^2+U(z)^2}}{I(z)}.
\label{eq:pol}
\end{equation}
The data term in Eq.~(\ref{eq:loss}) is the sum of the squared,
normalized residuals of $W(z)$ and $p(z)$ against their targets, on the
jet region $|z|>z_{\rm m}$, with $z_{\rm m}=3\,r_{\rm g}$ for the GRMHD
data; heights inside the mask carry no data.

The width is, by construction, blind to the pitch: scaling the toroidal
component at fixed energy leaves $|\mathbf{B}|^2$ and therefore $W(z)$
unchanged. The polarization, through the field angle $\chi$,
is sensitive to it. The two observables are thus complementary, and it is
their combination that constrains the geometry. We quantify both
statements in Sect.~\ref{sec:synth}.

\subsection{Optimization via automatic differentiation}
\label{sec:autodiff}

We minimize the functional in Eq.~(\ref{eq:loss}) by gradient descent.
The forward model and prior are implemented in JAX. This gives exact
gradients of $\mathcal{L}$ with respect to all $\sim\!3.5\times10^6$ field
degrees of freedom through automatic differentiation --- no finite-difference
approximation needed. The unit constraint on $\mathbf{S}$ is
imposed by normalizing at each step, and the non-negativity of $A$ by a
softplus reparametrization. We use the Adam optimizer with learning rate
$0.05$ for $400$ iterations, which is sufficient for the data residuals
to converge to the sub-percent level; convergence curves are shown in
Appendix~\ref{app:conv}. The field is initialized as a smooth, collimated
helical sheath with a small random perturbation; the initial guess has
$S_z\propto+\mathrm{sign}(z)$ and a toroidal component of fixed sign
(Appendix~\ref{app:prior}). As shown in Sect.~\ref{sec:degeneracy}, the
choice of these two signs is not innocuous.

\subsection{Domain, boundaries and axis}
\label{sec:bc}

We impose no explicit boundary conditions on the field. The spatial
derivatives entering the prior and the divergence penalty are evaluated by
centred finite differences in the interior and one-sided differences at
the faces of the box, with no field values prescribed or extrapolated
beyond the boundary. The domain is large enough that the jet field of
interest is well inside it, and the emission-weighted observables are
dominated by the interior, so the boundary does not drive the
reconstruction.

The jet axis is not treated as a special locus: the Cartesian lattice is
uniform and the axis is just the line of cells at $x=y=0$; no coordinate
singularity arises and no separate regularization is applied there. The
only region excluded from the data term is the band $|z|\le z_{\rm m}$ around
the disc midplane.

\subsection{Computational cost}
\label{sec:cost}

Each reconstruction on the $96^3$ lattice optimizes $\sim\!3.5\times10^6$
degrees of freedom (three direction components plus one amplitude per cell)
through $400$ iterations. One automatic-differentiation pass
per iteration evaluates the forward model, the prior, and their exact
gradients; this is compiled once and reused. A $96^3$ reconstruction takes
$1.5$--$3$\,min on a laptop or workstation CPU. The cost and the memory scale
linearly with the number of cells: the funnel lattice used in
Sect.~\ref{sec:sign} ($128^2\times96$ cells) costs about $1.7$ times more,
while a uniform $256^3$ cube would cost about $19$ times more in time and in
memory, beyond what is practical on a workstation for a library of hundreds
of snapshots. Uniformly refining the whole cube is therefore not the
economical way to resolve the jet; refining only the funnel is
(Sect.~\ref{sec:extraction}).

\subsection{Discrete degeneracies: formal statement}
\label{sec:degeneracy}

The construction above has two exact symmetries.

\paragraph{Proposition 1 (field reversal).} \emph{If the data term depends on
the field only through $W$ and $p$ (Eqs.~\ref{eq:width}--\ref{eq:pol}) and
the prior has $h=0$, then $\mathcal{L}[-\mathbf{B}]=\mathcal{L}[\mathbf{B}]$.}

\paragraph{Proof.} Under $\mathbf{B}\to-\mathbf{B}$ the emissivity
$\varepsilon\propto|\mathbf{B}|^2$ is unchanged, so $I$ and $W$ are
unchanged. The plane-of-sky angle transforms as
$\chi\to\chi+\pi$, hence $2\chi\to2\chi+2\pi$ and both $\cos 2\chi$ and
$\sin 2\chi$ are unchanged; $Q$, $U$ and $p$ are therefore invariant. Every
term of $\mathcal{H}$ with $h=0$ depends on $\mathbf{S}$ through $|S_z|$ or
squared gradients, and $\mathcal{L}_{\rm div}$ is quadratic in $\mathbf{B}$.
$\square$

\paragraph{Proposition 2 (mirror through the sky plane).} \emph{Let
$\mathcal{T}$ map a field to its mirror image through the plane of the sky,
\begin{equation}
(\mathcal{T}\mathbf{B})(x,y,z)=\bigl(B_x,\,-B_y,\,B_z\bigr)(x,-y,z).
\label{eq:mirror}
\end{equation}
Under the same assumptions, $\mathcal{L}[\mathcal{T}\mathbf{B}]=\mathcal{L}[\mathbf{B}]$.}

\paragraph{Proof.} $\varepsilon$ and $\chi$ at $(x,y,z)$ take the values they
had at $(x,-y,z)$, so every line-of-sight integral in
Eqs.~(\ref{eq:width})--(\ref{eq:pol}) is unchanged. The terms of
$\mathcal{H}$ depend on $|S_z|$, on $R=(x^2+y^2)^{1/2}$ and on squared
gradients, all invariant under the reflection; and
$\nabla\!\cdot\!(\mathcal{T}\mathbf{B})=(\nabla\!\cdot\!\mathbf{B})(x,-y,z)$,
so $\mathcal{L}_{\rm div}$ is invariant. On a lattice symmetric under
$y\to-y$ the statement is exact. $\square$

\smallskip
For an axisymmetric field, $\mathcal{T}$ acts on cylindrical components as
$(B_R,B_\phi,B_z)\to(B_R,-B_\phi,B_z)$: it reverses the handedness of the
helix at fixed poloidal field. The cost is therefore invariant under the
group $G=\{1,-1,\mathcal{T},-\mathcal{T}\}$, whose elements reverse, respectively,
nothing, both senses, the toroidal sense only, and the poloidal sense only
($-\mathcal{T}:(B_R,B_\phi,B_z)\to(-B_R,B_\phi,-B_z)$). We verified both
propositions numerically on synthetic and on randomly perturbed,
non-axisymmetric fields: $W$ and $p$ agree to $\sim\!10^{-15}$
(Fig.~\ref{fig:rm}).

\paragraph{Corollary (the initial guess selects the image).} Because
$\mathcal{L}$ is invariant under $G$ and each element of $G$ acts linearly
and isometrically on the parameters, gradient descent and Adam commute with
$G$: starting from $g\,\mathbf{B}_0$ the optimizer follows the image under
$g$ of the trajectory started from $\mathbf{B}_0$. Which of the four
images the reconstruction ends in is therefore fixed by the initial guess,
not by the data (up to the small random part of the initialization, which
is not symmetric). We confirm this directly in Sect.~\ref{sec:synth_sym}.

\smallskip
Stated in observational terms: from the width and polarization of a
side-on jet one can reconstruct the unsigned geometry of its field --- the
collimation, the magnitude of the pitch, whether field lines open or
close --- but neither whether the field threads the black hole inward or
outward, nor whether the helix is right- or left-handed. Both are
choices made by the reconstruction, not measurements.

\paragraph{Proposition 3 (rotation measure, side-on).} \emph{Let the data
also include the rotation-measure map $\mathrm{RM}(x,z)=\int n_e B_y\,dy$,
with $n_e$ even in $\mathbf{B}$. Then $\mathrm{RM}$ changes sign under $-1$
and under $\mathcal{T}$, and is invariant under $-\mathcal{T}$.}

\smallskip
The proof is immediate from Eq.~(\ref{eq:mirror}). A rotation-measure term
therefore reduces the degeneracy from $G$ to $\{1,-\mathcal{T}\}$: it fixes the
sign of the toroidal field but not that of the poloidal field. The
reason is geometric. For a side-on jet the line-of-sight component $B_y$
is supplied by $B_\phi$ (and by the small radial component), while $B_z$
is perpendicular to the line of sight and does not enter the rotation
measure at all. We test this in Sects.~\ref{sec:synth_sym} and
\ref{sec:sign_rm}. Proposition~3 concerns the RM map, i.e.\ an
external Faraday screen. When the emitting plasma itself rotates the
polarization and the polarization is resolved at several wavelengths, the
remaining degeneracy is also lifted, except for configurations that are
symmetric in depth (Proposition~4, Sect.~\ref{sec:sign_ftomo}).

\section{GRMHD simulations and data}
\label{sec:data}

\subsection{Simulation library}
\label{sec:library}

Our testbed is the public GRMHD library of \citet{dhruv2025}, computed with
the code \texttt{KHARMA} \citep{prather2024}, an independent, performance-portable
C++17 code that descends from \texttt{iharm3d} \citep{gammie2003}. The library
covers the two canonical accretion states
--- SANE and MAD \citep{narayan2003,tchekhovskoy11} --- at five spins,
$a_*=-15/16,\,-1/2,\,0,\,+1/2,\,+15/16$, for a total of ten models. Each run
extends to $3\times10^4\,GM\,c^{-3}$. For each model we use the last
$N_{\rm s}=10$ snapshots of the run, at a cadence of $5\,GM\,c^{-3}$, spanning
$t=25100$--$25145\,GM\,c^{-3}$ ($25215$--$25260$ for MAD $a_*=+0.94$) in the
quasi-steady state (Table~\ref{tab:models}). Each snapshot is reconstructed
independently; no time averaging of the field is performed. The statistics we
quote are means and standard deviations over the snapshots of a model. Over
this $45\,GM\,c^{-3}$ window they reflect the short-term variability of each
flow, not its long-term evolution. Throughout we quote the two highest-spin values as
$a_*=\pm0.94$ for brevity.

The two states differ in the degree of magnetic flux threading the
horizon. MAD flows build up a
dynamically important, ordered poloidal field and drive powerful,
well-collimated jets, most efficiently at high prograde spin
\citep{tchekhovskoy11,mckinney12}. SANE flows are more turbulent and
their field is less ordered. A method that recovers field geometry should
behave differently, and traceably, across the two.

\subsection{From the simulation grid to a Cartesian field cube}
\label{sec:extraction}

The simulations use the ``funky'' modified Kerr--Schild (FMKS) coordinates
$(X_1,X_2,X_3)$ of \texttt{iharm3d}, with $r=e^{X_1}$, $\phi=X_3$ and
$\theta(X_1,X_2)$ a smooth map that concentrates resolution towards the disc
midplane and, close to the black hole, widens the cells near the poles. The
native resolution is $288\times128\times128$. The dumps store the
contravariant field components $B^{i}$ in these coordinates. We convert
them to Kerr--Schild components with the Jacobian of the map,
\begin{equation}
B^r=r\,B^{1},\qquad
B^\theta=\frac{\partial\theta}{\partial X_1}B^{1}+\frac{\partial\theta}{\partial X_2}B^{2},\qquad
B^\phi=B^{3},
\end{equation}
normalize them by the diagonal of the Kerr--Schild metric,
$\hat B_r=\sqrt{g_{rr}}\,B^r$, $\hat B_\theta=\sqrt{g_{\theta\theta}}\,B^\theta$,
$\hat B_\phi=\sqrt{g_{\phi\phi}}\,B^\phi$, and finally project them on
Cartesian axes aligned with the spin axis. The magnetization is computed
from the fluid-frame field, $\sigma=b^2/\rho$, with $b^\mu$ built from
$B^i$, the fluid four-velocity and the Kerr--Schild metric; the jet is the
region $\sigma>1$ \citep{eht2021_viii}.

The Cartesian field is obtained by linear interpolation in the native
coordinates, after mapping each Cartesian cell centre to $(X_1,X_2,X_3)$. We
build two kinds of cube. In the \emph{mean-field} cube, used for the main
results, the native components are first averaged over $X_3$ at fixed
$(X_1,X_2)$. The poloidal field is the azimuthal mean of $(B^r,B^\theta)$, and
the toroidal field is the azimuthal mean of $B^\phi$, assigned along the local
$\hat{\mathbf{e}}_\phi$ at every Cartesian cell. Their combination is the
axisymmetric mean field of the snapshot. No winding is imposed: the
toroidal field keeps the sign and amplitude of the simulation at each
$(r,\theta)$. In the
\emph{full} cube no average is taken and the native field is interpolated in
all three coordinates. Both are sampled either on the $96^3$ cube of
half-width $30\,r_{\rm g}$ (spacing $0.63\,r_{\rm g}$) or on a funnel lattice
of $128\times128\times96$ cells covering $|x|,|y|\le10\,r_{\rm g}$ and
$|z|\le30\,r_{\rm g}$ (transverse spacing $0.16\,r_{\rm g}$).

The native grid itself limits the near-axis resolution. Close to the poles the
FMKS map widens the $\theta$ cells, so at
$r=3$, $10$ and $30\,r_{\rm g}$ the first polar cell has a transverse size of
$0.36$--$0.41$, $0.84$--$0.94$ and $1.98$--$2.15\,r_{\rm g}$ across the library
(Table~\ref{tab:native}, Appendix~\ref{app:models}). The wide MAD funnels
($\sigma>1$ up to $0.6$--$0.7$\,rad from the axis at $30\,r_{\rm g}$) are still spanned
by $12$--$64$ native $\theta$ cells. The SANE funnels are much narrower: at
$30\,r_{\rm g}$ they span at most eight native cells, and in two of the five
SANE snapshots inspected there is no $\sigma>1$ region on the polar cells at
that radius. Beyond $r\approx7\,r_{\rm g}$ the Cartesian spacing of the $96^3$
cube is finer than the native polar cells, and the spacing of the funnel
lattice is finer everywhere in the jet. Resampling therefore does not
coarse-grain the transverse structure of the jet. The smallest near-axis
structure available to any reconstruction, however sampled, is set by the
native polar resolution of the simulations, most severely for the SANE jets.

\subsection{Observables}
\label{sec:obs}

From each field cube, restricted to the jet ($\sigma>1$), we compute the two
observables of Sect.~\ref{sec:forward}, $W(z)$ and $p(z)$, using the same
forward model that the inverse method inverts, and mask
$|z|\le z_{\rm m}=3\,r_{\rm g}$. Snapshots in which the $\sigma>1$ funnel
contains fewer than $200$ cells, which happens when the jet is transiently
weak, would be excluded; none of the $100$ snapshots used falls in this
category. The reconstruction is then run on
$W(z)$ and $p(z)$ alone, with the true field retained only to score the
result.

\section{Validation on synthetic jets}
\label{sec:synth}

Before applying the method to GRMHD data, we test it on synthetic jets with
exactly known fields. This isolates the behaviour of the method from the
complications of the simulations and lets us check, term by term, that the
reconstruction does what the forward model promises.

\subsection{Setup and alignment metrics}
\label{sec:synth_setup}

The synthetic jet is a collimating helical field on a $96^3$ lattice of
half-width $10$ length units (one unit corresponds to $3\,r_{\rm g}$ if the
lattice is identified with the GRMHD cube). The field is confined to a sheath
around a parabolic boundary, $R_{\rm j}\propto |z|^{k}$ with $k=0.55$, with
$B_z\propto\mathrm{sign}(z)$ and a toroidal component of fixed sign and
constant pitch $\mu=|B_\phi|/|B_z|$. From this known field we generate $W(z)$ and $p(z)$ through
the forward model of Sect.~\ref{sec:forward} and reconstruct the field from
those observables alone; for the synthetic jets the mask is
$z_{\rm m}=0.6$ units.

We score a reconstruction cell by cell. At each lattice cell $i$ inside the
jet we take the cosine of the angle between the \emph{true} and the
\emph{reconstructed} magnetic-field vectors at that cell,
\begin{equation}
\cos\vartheta_i=\frac{\mathbf{B}_{\rm true}(\mathbf{x}_i)\cdot\mathbf{B}_{\rm rec}(\mathbf{x}_i)}
{|\mathbf{B}_{\rm true}(\mathbf{x}_i)|\,|\mathbf{B}_{\rm rec}(\mathbf{x}_i)|}.
\label{eq:cos}
\end{equation}
The angle $\vartheta_i$ is an angle between two field vectors at the same
point; it involves neither the black-hole spin axis nor the line of sight.
We average over the jet cells, defined as $|z|>z_{\rm m}$ and
$|\mathbf{B}_{\rm true}|>0.2\max|\mathbf{B}_{\rm true}|$ (the maximum being
taken over $|z|>z_{\rm m}$), with equal weight per cell. The signed mean
$\langle\cos\rangle$ equals $+1$ only for a reconstruction that is correct in
both orientation and sense, and $0$ for random directions. The unsigned mean
$\langle|\cos|\rangle$ is invariant under $\mathbf{B}_{\rm rec}\to-\mathbf{B}_{\rm rec}$
and measures orientation irrespective of sense; for random directions it
equals $1/2$ (the mean of $|\cos|$ over the sphere). Neither is invariant
under the mirror $\mathcal{T}$: a reconstruction that lands on the mirror
image of the truth scores low even though it fits the data equally well.
We therefore also quote $\langle|\cos|\rangle_G$, the larger of the scores of
$\mathbf{B}_{\rm rec}$ and of $\mathcal{T}\mathbf{B}_{\rm rec}$, which is
blind to everything the data cannot determine.

\subsection{Width is blind to pitch; polarization is not}
\label{sec:synth_degen}

\begin{figure}
\centering
\includegraphics[width=\columnwidth]{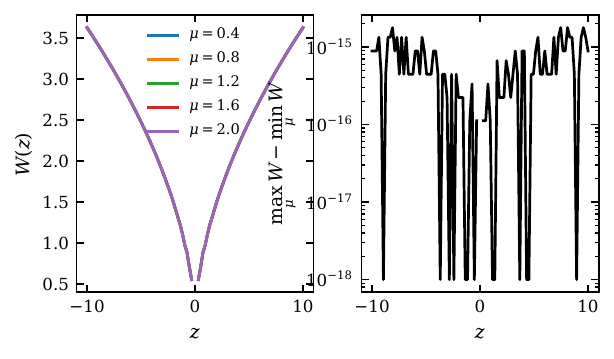}
\caption{The jet width is blind to the magnetic pitch. \emph{Left:}
$W(z)$ for five values of the pitch $\mu$; the curves overlap.
\emph{Right:} the maximum spread of $W$ across $\mu$ is at the level of
$10^{-15}$, i.e.\ machine precision.}
\label{fig:degen}
\end{figure}

\begin{figure}
\centering
\includegraphics[width=\columnwidth]{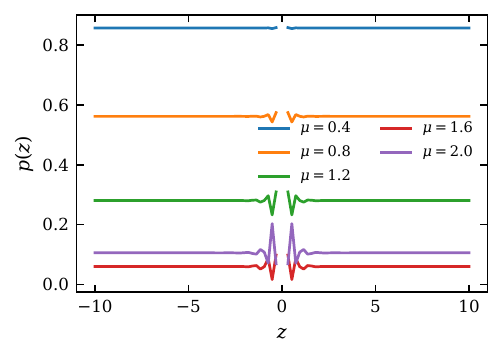}
\caption{The polarization degree $p(z)$ for the same five values of the
pitch. Unlike the width, $p(z)$ responds strongly to the pitch, varying
by more than a factor of two.}
\label{fig:pol}
\end{figure}

The synthetic jet makes the complementary roles of the two observables
explicit. When we scan the pitch over $\mu\in[0.4,2.0]$ at fixed magnetic
energy, the jet width $W(z)$ does not change: variations are at the
$10^{-15}$ level, i.e.\ machine precision (Fig.~\ref{fig:degen}). The width
is set by the energy distribution, which is held fixed. The
polarization degree $p(z)$, by contrast, changes by more than a factor of
two over the same range (Fig.~\ref{fig:pol}), because it depends on the
projected field angle. Width constrains the envelope; polarization
constrains the winding. Neither alone fixes the geometry.

\subsection{Reconstruction of a known field}
\label{sec:synth_recon}

\begin{figure*}
\centering
\includegraphics[width=\textwidth]{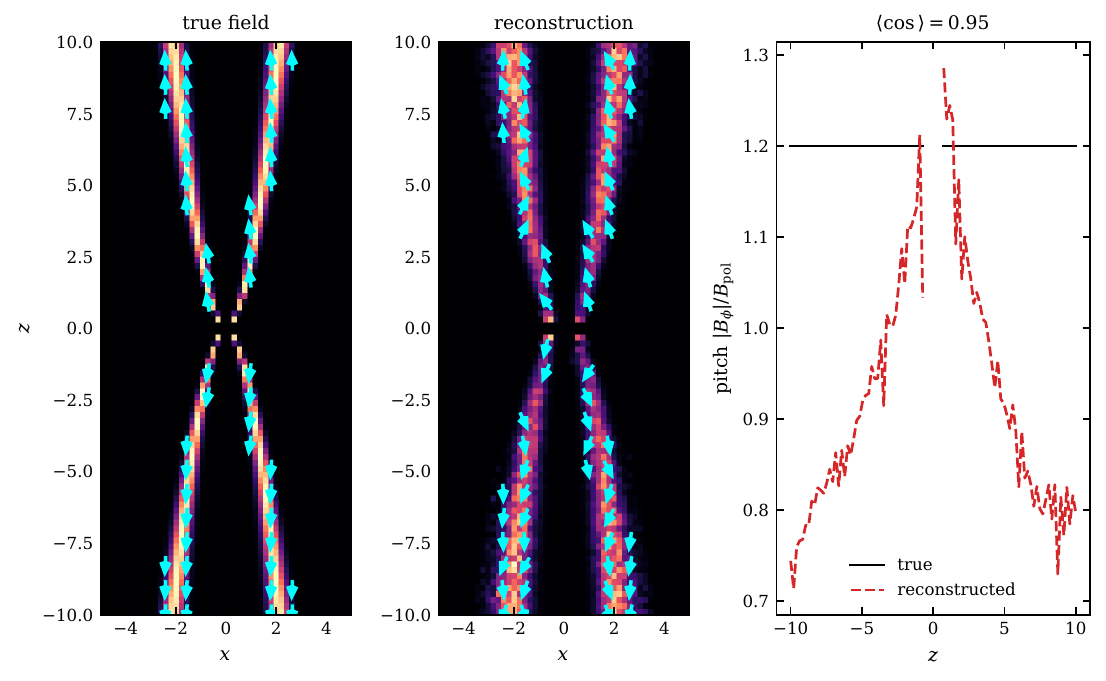}
\caption{Reconstruction of a synthetic jet with known truth ($\mu=1.2$).
\emph{Left:} the true field in the $y=0$ plane (colour: $|\mathbf{B}|^2$;
arrows: in-plane field direction). \emph{Centre:} the H-MOG reconstruction
from $W$ and $p$ only, starting from an initial guess with the same senses as
the truth. \emph{Right:} the true and reconstructed pitch profiles
$\mu(z)$. The direction field is recovered at $\langle\cos\rangle=0.95$.}
\label{fig:synthrec}
\end{figure*}

Starting from the default initial guess, whose senses coincide with those of
the synthetic truth, H-MOG recovers the field with
$\langle\cos\rangle=\langle|\cos|\rangle=0.95$ over the jet body for
$\mu=1.2$ (Fig.~\ref{fig:synthrec}). The
width is reproduced to $0.06\%$ and the polarization to better than
$0.01\%$. The local pitch, however, is not fixed by the data along the whole
jet (Fig.~\ref{fig:synthrec}, right). It matches the true value near the base
($1.2$ at $|z|\simeq1$) and falls to $\simeq0.8$ ($-33\%$) at $|z|\simeq10$.
The polarization $p(z)$ is a single number per height, so it constrains the
slice-integrated projected field angle, which can be matched by different
combinations of pitch and radial (opening) field. Within that freedom the
prior decides. The external-pressure term pulls the opening angle towards a
target $\theta_{\rm t}(z)$ that decreases with height, and so lowers the pitch
there. Without it ($\alpha=0$) the pitch is instead biased high, reaching
$1.3$--$1.55$ at $|z|\ge5$, while the direction-field alignment improves to
$0.985$. The pitch profile is therefore the most prior-dependent part of the
reconstructed geometry. The direction field as a whole is robust
(Sect.~\ref{sec:prior_sens}).

\subsection{Skill across the pitch range}
\label{sec:synth_pitch}

\begin{figure}
\centering
\includegraphics[width=\columnwidth]{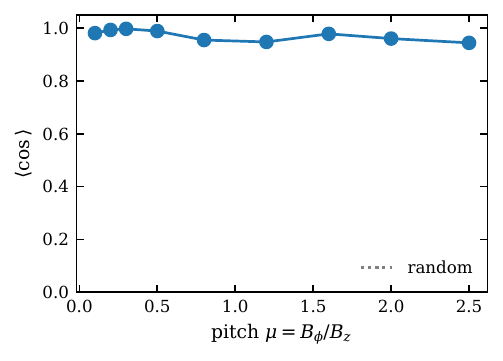}
\caption{Reconstruction skill $\langle\cos\rangle$ as a function of the
jet pitch $\mu$, for synthetic jets with known truth and an initial guess
with the same senses as the truth. The alignment remains close to unity
across the whole range, far above the random expectation (dotted).}
\label{fig:skillpitch}
\end{figure}

The result above is not specific to one pitch. Repeating the
reconstruction for synthetic jets with $\mu$ from $0.1$ to $2.5$, the
alignment stays between $0.94$ and $1.00$ throughout, with shallow dips at
$\mu\simeq1$ and $\mu=2.5$ (Fig.~\ref{fig:skillpitch}). In
these runs the initial guess has the same senses as the truth, so
$\langle\cos\rangle=\langle|\cos|\rangle$.

\subsection{Robustness to observational noise}
\label{sec:synth_noise}

\begin{figure}
\centering
\includegraphics[width=\columnwidth]{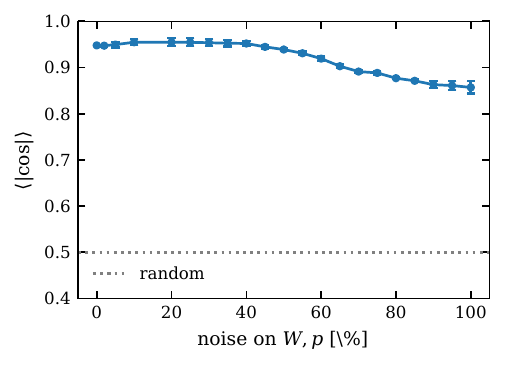}
\caption{Robustness of the reconstructed geometry to observational
noise. The unsigned alignment $\langle|\cos|\rangle$ remains close to
$0.95$ up to $40\%$ noise on $W$ and $p$, and degrades only to $0.86$ at
$100\%$, never approaching the random value of $0.5$ (dotted).}
\label{fig:noise}
\end{figure}

We test graceful degradation by adding Gaussian
noise to $W(z)$ and $p(z)$ at levels from $0$ to $100\%$ of the signal,
then repeating the reconstruction (Fig.~\ref{fig:noise}). The geometry
holds up well. $\langle|\cos|\rangle$ stays near $0.95$ up to
$\sim\!40\%$ noise, then declines gently to $\simeq0.86$ at $100\%$.
It never approaches the random value of $0.5$. The Hamiltonian
prior acts as a regularizer: it absorbs noise while preserving geometry.

\subsection{Sensitivity to the prior}
\label{sec:prior_sens}

A variational method with a physical prior raises an obvious worry: how
much of the result is data, and how much is prior? We vary the three
weights --- global $\alpha_{\rm prior}$, external-pressure $\alpha$, and
shear $\beta$ --- one at a time around their defaults.

\begin{figure*}
\centering
\includegraphics[width=\textwidth]{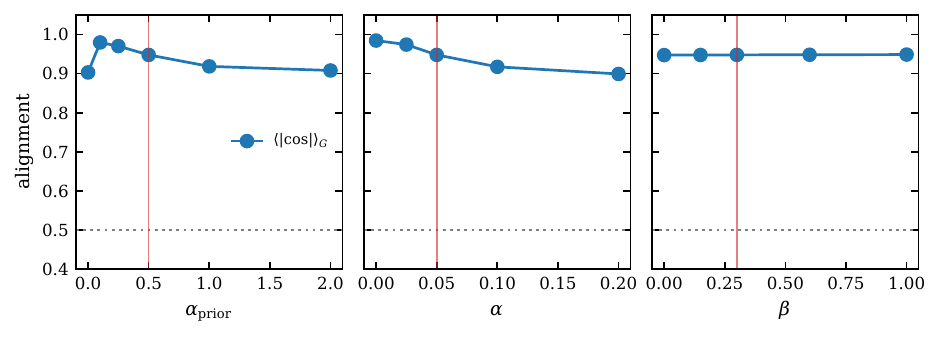}
\caption{Sensitivity of the synthetic reconstruction to the prior weights.
Each panel varies one weight around its default (vertical line): the
global prior weight $\alpha_{\rm prior}$, the external-pressure weight
$\alpha$, and the shear weight $\beta$. The dotted line marks the random
value $0.5$ of $\langle|\cos|\rangle$.}
\label{fig:priorsens}
\end{figure*}

The geometry is stable (Fig.~\ref{fig:priorsens}). The
alignment stays in the range $0.91$--$0.98$ as the global prior weight is
varied by a factor of twenty ($\alpha_{\rm prior}=0.1$ to $2.0$), and it
remains $0.90$ when the prior is switched off entirely. The
external-pressure weight has the largest effect: the alignment is $0.985$
without it and falls to $0.90$ at the largest value tested ($\alpha=0.2$, four
times the default), consistent with its influence on the pitch profile
(Sect.~\ref{sec:synth_recon}). The shear weight moves the alignment by less
than $0.002$ across its full range; with our normalization the shear window
lies outside most of the funnel (Appendix~\ref{app:prior}), where little
emission resides. In every case the width and polarization are fitted to
better than $0.1\%$. Within the image
selected by the initial guess, the data fix the geometry and the prior
regularizes it without dictating it.

\subsection{The initial guess selects the sense and the handedness}
\label{sec:synth_sym}

\begin{figure}
\centering
\includegraphics[width=\columnwidth]{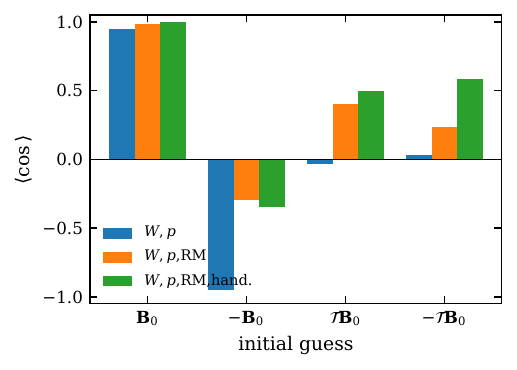}
\caption{Signed alignment $\langle\cos\rangle$ of synthetic reconstructions
started from the four images of the default initial guess, for three data
sets: $W,p$; $W,p$ and RM; $W,p$, RM and a handedness prior. With $W,p$ the
result follows the initial guess. With the RM term the run started with the
wrong poloidal sense keeps it. With RM and handedness the lowest final cost is
reached by the true image (Table~\ref{tab:symtest}).}
\label{fig:symtest}
\end{figure}

The corollary of Sect.~\ref{sec:degeneracy} predicts that the four images
$\{\mathbf{B},-\mathbf{B},\mathcal{T}\mathbf{B},-\mathcal{T}\mathbf{B}\}$ of
the truth fit the data equally well, and that the reconstruction lands in
the image selected by the initial guess. We test this by starting the
synthetic reconstruction from the four images of the default initial guess
(Fig.~\ref{fig:symtest}, Table~\ref{tab:symtest}). With $W$ and $p$ only, the four
runs end at the same cost, to $0.2\%$, and each converges to the image of its
initial guess. Starting from $-\mathbf{B}_0$ gives $\langle\cos\rangle=-0.95$
with unchanged $\langle|\cos|\rangle$. Starting from the mirror images gives
$\langle|\cos|\rangle=0.09$, i.e.\ a helix of the opposite handedness, while
$\langle|\cos|\rangle_G=0.95$: the reconstruction is the mirror image of the truth,
recovered as accurately as the truth itself.

Adding a rotation-measure term ($n_e\propto|\mathbf{B}|^2$, weight $2$) does
what Proposition~3 predicts. Started from $-\mathcal{T}\mathbf{B}_0$ (poloidal sense
wrong, toroidal sense right), the reconstruction fits the RM map with a
correlation coefficient of $1.00$ and ends in the image $-\mathcal{T}\mathbf{B}$,
with $\langle\cos\rangle=0.24$ and $\langle|\cos|\rangle_G=0.985$, at the same cost
as the run started from $\mathbf{B}_0$ (to $0.1\%$). The same holds for every RM
weight from $0.5$ to $5$ and for an electron density fixed to the true one. Runs
started with the wrong toroidal sense ($\mathcal{T}\mathbf{B}_0$, $-\mathbf{B}_0$)
are pushed by the RM term away from their image, but within 400 iterations they
end in mixed states ($\langle\cos\rangle=0.40$ and $-0.29$) at a $12\%$ higher
cost. The RM thus removes two of the four images and leaves $\mathbf{B}$ and
$-\mathcal{T}\mathbf{B}$ exactly degenerate. The poloidal sense is recovered only when
the handedness is fixed as well (Sect.~\ref{sec:sign_route}).

\begin{table}
\caption{Synthetic reconstructions ($\mu=1.2$, $96^3$, 400 iterations) started from the
four images of the default initial guess $\mathbf{B}_0$. $\Delta\mathcal{L}$ is the
final cost relative to the lowest one obtained with the same data.}
\label{tab:symtest}
\centering
\begin{tabular}{l l c c c c}
\hline\hline
Data & Start & $\langle\cos\rangle$ & $\langle|\cos|\rangle$ & $\langle|\cos|\rangle_G$ & $\Delta\mathcal{L}$ \\
\hline
$W,p$ & $\mathbf{B}_0$             & $+0.95$ & $0.95$ & $0.95$ & $0.04\%$ \\
      & $-\mathbf{B}_0$            & $-0.95$ & $0.95$ & $0.95$ & $0.02\%$ \\
      & $\mathcal{T}\mathbf{B}_0$  & $-0.03$ & $0.09$ & $0.95$ & $0.16\%$ \\
      & $-\mathcal{T}\mathbf{B}_0$ & $+0.03$ & $0.09$ & $0.95$ & $0$ \\
\hline
$W,p,$RM & $\mathbf{B}_0$             & $+0.99$ & $0.99$ & $0.99$ & $0$ \\
         & $-\mathbf{B}_0$            & $-0.29$ & $0.43$ & $0.45$ & $12\%$ \\
         & $\mathcal{T}\mathbf{B}_0$  & $+0.40$ & $0.45$ & $0.45$ & $12\%$ \\
         & $-\mathcal{T}\mathbf{B}_0$ & $+0.24$ & $0.24$ & $0.99$ & $0.1\%$ \\
\hline
$W,p,$RM, & $\mathbf{B}_0$             & $+1.00$ & $1.00$ & $1.00$ & $0$ \\
handed.   & $-\mathbf{B}_0$            & $-0.35$ & $0.82$ & $0.82$ & $13\%$ \\
          & $\mathcal{T}\mathbf{B}_0$  & $+0.50$ & $0.59$ & $0.59$ & $20\%$ \\
          & $-\mathcal{T}\mathbf{B}_0$ & $+0.59$ & $0.71$ & $0.71$ & $25\%$ \\
\hline
\end{tabular}
\end{table}

\subsection{Resolution does not lift the degeneracy}
\label{sec:synth_res}

\begin{figure}
\centering
\includegraphics[width=\columnwidth]{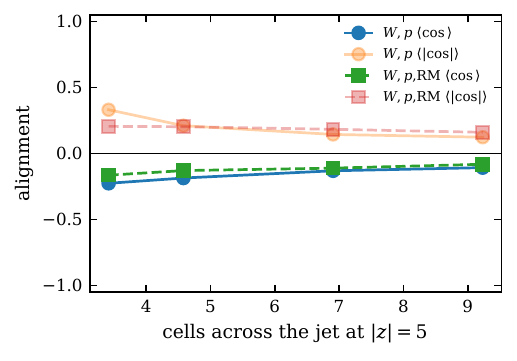}
\caption{Synthetic RM test at increasing lattice resolution for a narrow jet.
Squares: signed alignment with an RM term; circles: without. Starting from
the image with the wrong poloidal sense, the poloidal sense is not recovered
at any resolution.}
\label{fig:rmres}
\end{figure}

To separate symmetry from resolution, we repeat the RM test on a narrower
synthetic jet ($R_{\rm j}$ halved) on lattices with $N=48$, $64$, $96$ and
$128$, so that the jet diameter at $|z|=5$ units spans from
$\sim\!3$ to $\sim\!9$ cells. Starting in the image
$-\mathcal{T}\mathbf{B}_0$, the reconstruction remains in that image at every
resolution: with the RM term, $\langle\cos\rangle=-0.16$, $-0.13$, $-0.11$
and $-0.08$ for $N=48$, $64$, $96$ and $128$ (Fig.~\ref{fig:rmres}), while the
RM map is fitted with a correlation coefficient of $1.00$ at every resolution.
Finer sampling does not select the image.

\section{Application to GRMHD jets}
\label{sec:geom}

We now run H-MOG on the ten GRMHD models of Sect.~\ref{sec:data}. For
each snapshot we reconstruct the field from the projected $W(z)$ and $p(z)$
of the mean-field cube, starting from the default initial guess, and score
against the simulated field with the metrics of Sect.~\ref{sec:synth_setup}.

\subsection{Geometry is recovered up to the mirror image}
\label{sec:geom_overall}

\begin{figure}
\centering
\includegraphics[width=\columnwidth]{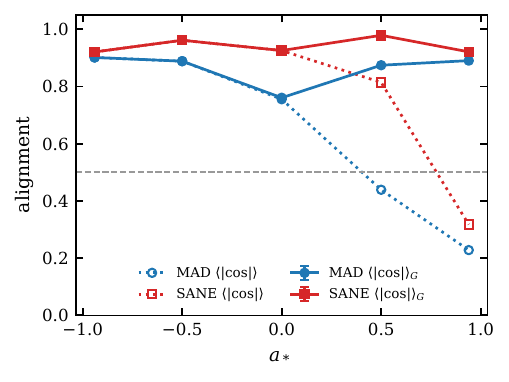}
\caption{Reconstruction accuracy across the GRMHD library against black-hole
spin. Filled symbols: handedness-blind score $\langle|\cos|\rangle_G$, with the
snapshot-to-snapshot scatter. Open symbols: $\langle|\cos|\rangle$ of the
reconstruction itself, which is low for the prograde models because the
reconstruction is the mirror image of the simulated field. Dashed line: random
value $0.5$.}
\label{fig:skillspin}
\end{figure}

Table~\ref{tab:grmhd} and Fig.~\ref{fig:skillspin} summarize the
reconstructions. The handedness-blind score $\langle|\cos|\rangle_G$ lies
between $0.76$ and $0.90$ for the MAD models and between $0.92$ and $0.98$ for
the SANE models, well above the random value $0.5$. The scatter between
snapshots is below $0.01$. The unsigned score of the reconstruction itself,
$\langle|\cos|\rangle$, tells the story of Proposition~2. It equals
$\langle|\cos|\rangle_G$ for the retrograde models ($0.89$--$0.96$) and drops to
$0.23$--$0.44$ (MAD) and $0.32$--$0.81$ (SANE) for the prograde ones, where the
reconstruction is the mirror image of the simulated field. Its axis
correlation then turns negative, as expected for a helix of the opposite
handedness. The reason is measured directly in the simulations
(Sect.~\ref{sec:sign_diag}). The handedness of the jet helix follows the sign
of the spin, and the default initial guess has the handedness of the
retrograde models. In every model the widths and polarizations are fitted to
better than $3\%$ ($35\%$ in $p$ for SANE $a_*=+0.94$), so the mirror image
is not a poor fit: it is an equally good one.

The two accretion states differ in the geometry. The SANE jets are recovered
better than the MAD ones, and the MAD model with $a_*=0$ is the hardest case
($\langle|\cos|\rangle_G=0.76$). There the handedness differs between the two
jets (Sect.~\ref{sec:sign_diag}), so that neither the reconstruction nor its
mirror image matches both. In MAD flows the $\sigma>1$ region is also much
wider than in SANE flows (half-opening angle $0.6$--$0.7$\,rad at $30\,r_{\rm g}$
against at most $0.4$\,rad), and it is likely to include more of the less
ordered field at the funnel edge.

The scores refer to the jet cells defined in Sect.~\ref{sec:synth_setup}, i.e.\
those with $|\mathbf{B}|>0.2\max|\mathbf{B}|$ at $|z|>3\,r_{\rm g}$. These are
about $9\%$ of the $\sigma>1$ cells but carry about half of the emission, and
they lie mostly at $|z|\lesssim10\,r_{\rm g}$. Over all $\sigma>1$ cells the
handedness-blind score of the representative snapshots is lower: $0.79$ for MAD
$a_*=+0.94$ (against $0.89$) and $0.82$ for MAD $a_*=-0.94$ (against $0.90$).
The outer, fainter parts of the funnel, which contribute little to $W$ and $p$,
are reconstructed less well. The method holds up across both accretion states
and the full spin range, on fields that were not used to design the prior or to
choose its weights.

\begin{table}
\caption{H-MOG reconstruction of the GRMHD library (mean-field cube, $96^3$,
default initial guess, $N_{\rm s}=10$ snapshots per model):
handedness-blind alignment $\langle|\cos|\rangle_G$ and unsigned alignment
$\langle|\cos|\rangle$, with their snapshot scatter. The full statistics are in
Table~\ref{tab:fulldata}.}
\label{tab:grmhd}
\centering
\begin{tabular}{l c c c}
\hline\hline
State & $a_*$ & $\langle|\cos|\rangle_G$ & $\langle|\cos|\rangle$ \\
\hline
MAD  & $-0.94$ & $0.901\pm0.001$ & $0.901\pm0.001$ \\
MAD  & $-0.50$ & $0.889\pm0.005$ & $0.889\pm0.005$ \\
MAD  & $ 0.00$ & $0.761\pm0.000$ & $0.755\pm0.001$ \\
MAD  & $+0.50$ & $0.874\pm0.000$ & $0.439\pm0.001$ \\
MAD  & $+0.94$ & $0.890\pm0.001$ & $0.228\pm0.002$ \\
\hline
SANE & $-0.94$ & $0.921\pm0.002$ & $0.921\pm0.002$ \\
SANE & $-0.50$ & $0.962\pm0.001$ & $0.962\pm0.001$ \\
SANE & $ 0.00$ & $0.926\pm0.002$ & $0.926\pm0.002$ \\
SANE & $+0.50$ & $0.979\pm0.001$ & $0.814\pm0.007$ \\
SANE & $+0.94$ & $0.921\pm0.006$ & $0.318\pm0.023$ \\
\hline
\end{tabular}
\end{table}

\subsection{Sensitivity to the prior on GRMHD data}
\label{sec:geom_prior}

\begin{figure*}
\centering
\includegraphics[width=\textwidth]{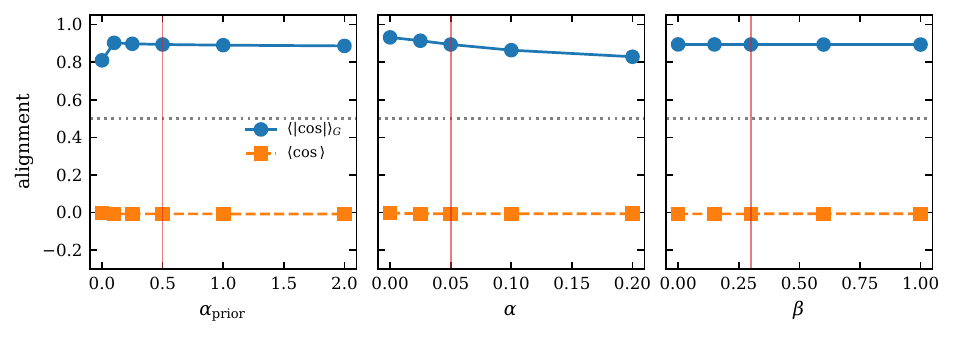}
\caption{As Fig.~\ref{fig:priorsens}, for the representative GRMHD snapshot
(MAD, $a_*=+0.94$). Circles: handedness-blind alignment
$\langle|\cos|\rangle_G$; squares: signed alignment $\langle\cos\rangle$.}
\label{fig:priorsens_grmhd}
\end{figure*}

The prior sweep of Sect.~\ref{sec:prior_sens}, repeated on the representative
GRMHD snapshot (MAD, $a_*=+0.94$, $t=25260\,GM\,c^{-3}$), gives the same
picture (Fig.~\ref{fig:priorsens_grmhd}). With any non-zero global weight
($\alpha_{\rm prior}=0.1$--$2$) the handedness-blind alignment stays between
$0.886$ and $0.902$, and it is $0.81$ without the prior. The external-pressure
weight matters more here than for the synthetic jet: $\langle|\cos|\rangle_G$
falls from $0.93$ at $\alpha=0$ to $0.83$ at $\alpha=0.2$. The target
inclination of $\mathcal{H}_{\rm ext}$ ($\theta_{\rm t}\simeq18^\circ$ at
$|z|=10\,r_{\rm g}$) is much smaller than that of the MAD field, whose axial
fraction $B_z/|\mathbf{B}|\simeq0.5$ corresponds to $\simeq60^\circ$, so the
term pulls against the data. The shear weight changes it by less than $10^{-3}$. Throughout,
$\langle\cos\rangle$ stays between $-0.007$ and $-0.001$, and $W$ and $p$ are
fitted to better than $0.5\%$. Since every term of the prior is invariant under
$G$, no choice of weights can move the reconstruction from one image to
another; the signed alignment is set by the initial guess.

\subsection{Visual validation}
\label{sec:geom_visual}

\begin{figure*}
\centering
\includegraphics[width=\textwidth]{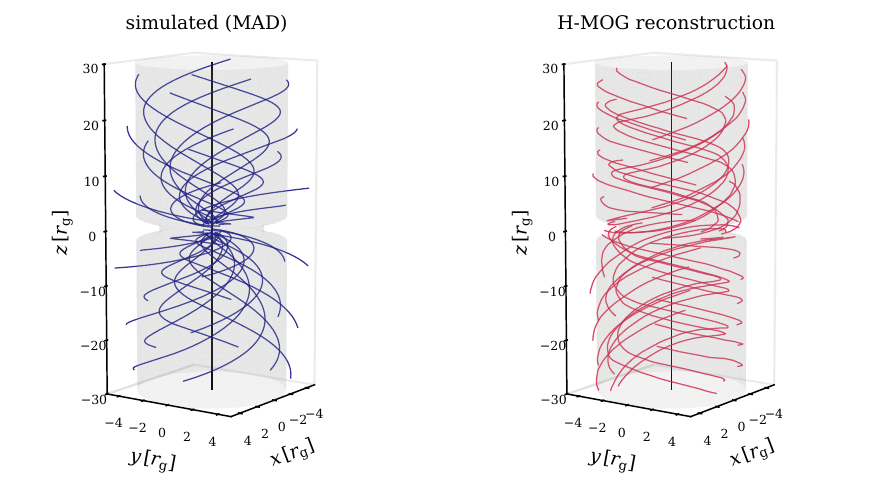}
\caption{Three-dimensional magnetic field lines of a representative GRMHD
jet (MAD, $a_*=+0.94$, $t=25260\,GM\,c^{-3}$), within $|x|,|y|\le5\,r_{\rm g}$. \emph{Left:} the
simulated (mean) field. \emph{Right:} the H-MOG reconstruction from $W$ and
$p$, which is the mirror image of the simulated field (opposite handedness,
Proposition~2). The grey surface is the $\sigma=1$ funnel boundary and the vertical line
the spin axis. We do not show arrows on the field lines, since the sense of
the field is not constrained by the observables
(Sect.~\ref{sec:degeneracy}).}
\label{fig:fieldlines}
\end{figure*}

Figure~\ref{fig:fieldlines} shows field lines of a representative GRMHD
model, traced in three dimensions from the simulated and the reconstructed
fields, within $|x|,|y|\le5\,r_{\rm g}$. The translucent surface marks the
$\sigma=1$ boundary of the funnel in the simulation and the vertical line the
spin axis; they show the narrowing of the jet towards the black hole and the
winding of the field lines about the axis. The reconstruction reproduces the
collimation and the pitch of the winding, but its helix winds the other way:
for this prograde model it is the mirror image of the simulated field
(Sect.~\ref{sec:geom_overall}). The field lines carry no arrows, because the
sense of the field is not constrained by the data.

\section{Routes to the field sense}
\label{sec:sign}

On the GRMHD library the signed alignment tells a different story from the
unsigned one. While $\langle|\cos|\rangle_G=0.76$--$0.98$, the signed
alignment $\langle\cos\rangle$ lies between $-0.13$ and $+0.42$, and between
$-0.02$ and $+0.03$ for all MAD models; it never approaches $+1$. In this section we explain this value and test three
routes to the sense.

\subsection{The degeneracy is exact}
\label{sec:sign_theory}

\begin{figure}
\centering
\includegraphics[width=\columnwidth]{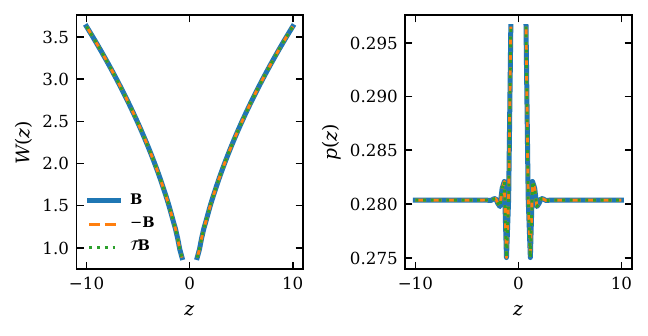}
\caption{The width $W(z)$ (left) and polarization $p(z)$ (right) computed
from a synthetic field $\mathbf{B}$, from its reverse $-\mathbf{B}$ and from
its mirror image $\mathcal{T}\mathbf{B}$ through the plane of the sky are
identical to machine precision.}
\label{fig:rm}
\end{figure}

Figure~\ref{fig:rm} illustrates Propositions~1 and 2. The width and
polarization profiles computed from a field, from its reverse and from its
mirror image coincide to machine precision. No term of the cost
functional can distinguish them.

\subsection{Where, and why, the sign is wrong}
\label{sec:sign_spatial}

\begin{figure}
\centering
\includegraphics[width=\columnwidth]{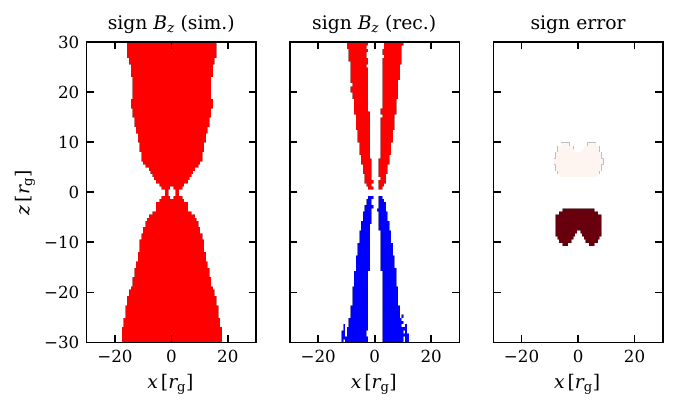}
\caption{Sign of $B_z$ in the slice $y=0$ for a representative GRMHD model
(MAD, $a_*=+0.94$, $t=25260\,GM\,c^{-3}$): in the simulated field (left) and in the
reconstruction (centre), and the cells where the two disagree (right,
scored jet cells only). The simulated poloidal field has the same direction in both
jets; the reconstruction inherits the sign pattern $\mathrm{sign}(z)$ of its
initial guess and is wrong in one hemisphere.}
\label{fig:signmap}
\end{figure}

The value of $\langle\cos\rangle$ follows from the corollary of
Sect.~\ref{sec:degeneracy}. The default initial guess has
$B_z\propto+\mathrm{sign}(z)$, i.e.\ a poloidal field pointing away from the
black hole in both jets, like a split monopole. In all ten simulations the
poloidal field instead keeps the same direction in both jets, threading the
black hole from one pole to the other, as a dipole does
(Fig.~\ref{fig:signmap}, left, and Sect.~\ref{sec:sign_diag}). The
reconstruction inherits the monopole-like sense of its initial guess, so it is
right in one jet and wrong in the other (Fig.~\ref{fig:signmap}). The sign error is therefore
hemispheric, not concentrated near the axis. $\langle\cos\rangle$ measures the
imbalance between the two hemispheres, not the quality of the reconstruction.
For the representative MAD $a_*=+0.94$
snapshot, the four sign variants of the default guess all give
$|\langle\cos\rangle|<0.01$. Their $\langle|\cos|\rangle$ is $0.22$ or $0.89$
depending on whether their handedness is opposite to or equal to that of the
simulation, while $\langle|\cos|\rangle_G=0.89$ in all four. A dipole-like
guess ($B_z\propto+1$ in both jets) gives $\langle\cos\rangle=+0.38$, and its
reverse gives $-0.38$. With a toroidal field of one sign in both jets, this
guess has the correct handedness in one jet only, and
$\langle|\cos|\rangle=0.52$. In every case the signed alignment is set by the
initial guess.

\subsection{Test 1: a rotation-measure term in the loss}
\label{sec:sign_rm}

The rotation measure, $\mathrm{RM}\propto\int n_e B_\parallel\,d\ell$, is
linear in the field and is the obvious candidate to break the degeneracy. We
add to the cost the term
$w_{\rm RM}\langle(\mathrm{RM}-\mathrm{RM}_{\rm obs})^2\rangle/\langle\mathrm{RM}_{\rm obs}^2\rangle$,
which matches the full projected map $\mathrm{RM}(x,z)$. For the electron density we use
two models. The first is $n_e\propto|\mathbf{B}|^2$, a simple proxy that ties
the electrons to the magnetic energy. The second is $n_e\propto\rho$ from the simulation,
with the caveat that in the $\sigma>1$ funnel the density is set by numerical
floors and is not a reliable tracer of the emitting electrons
\citep{eht2021_viii,moscibrodzka2017}.

\begin{figure}
\centering
\includegraphics[width=\columnwidth]{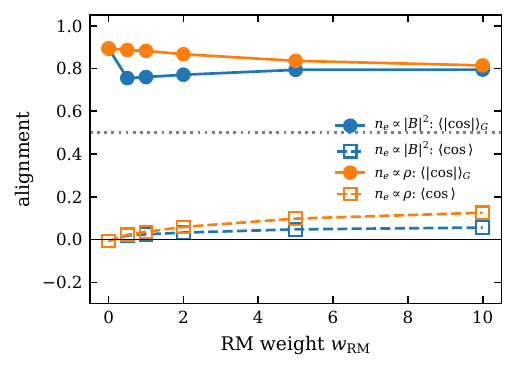}
\caption{Adding a rotation-measure term to the loss, for the representative
GRMHD snapshot (MAD, $a_*=+0.94$), with $n_e\propto|\mathbf{B}|^2$ and
$n_e\propto\rho$: handedness-blind alignment $\langle|\cos|\rangle_G$ (circles) and
signed alignment $\langle\cos\rangle$ (squares) against the weight of the RM
term.
The poloidal sense is not retrieved at any weight.}
\label{fig:rmsweep}
\end{figure}

On the GRMHD data the RM term does not recover the poloidal sense. For the
representative snapshot (MAD, $a_*=+0.94$) and RM weights from $0.5$ to $10$,
$\langle\cos\rangle$ stays between $0.02$ and $0.06$ with
$n_e\propto|\mathbf{B}|^2$, and below $0.13$ with $n_e\propto\rho$. Meanwhile the
handedness-blind geometry degrades from $0.89$ to $0.75$--$0.79$ and to
$0.81$--$0.89$, respectively (Fig.~\ref{fig:rmsweep}). The same holds on the
funnel lattice, whose transverse spacing ($0.16\,r_{\rm g}$) is four times finer.
For three snapshots each of the MAD and SANE $a_*=+0.94$ models we obtain
$\langle\cos\rangle=0.005$ and $0.03$ ($n_e\propto|\mathbf{B}|^2$) and $0.011$
and $0.10$ ($n_e\propto\rho$). For the SANE model the $|\mathbf{B}|^2$ proxy
conflicts with the data, and the RM term then spoils the fit of $W$ and $p$
(residuals of $28\%$ and $40$--$100\%$).

This is the behaviour Proposition~3 predicts, and the synthetic tests of
Sects.~\ref{sec:synth_sym} and \ref{sec:synth_res} show that it does not
depend on resolution. For a side-on jet the axial field is perpendicular to
the line of sight and does not enter the rotation measure. The RM therefore
constrains the toroidal field and leaves the image $-\mathcal{T}\mathbf{B}$,
which has the reversed poloidal field, as good a solution as the truth. The
failure is a symmetry of the observables, not a consequence of the azimuthal
averaging or of the sampling of the field cube.

\subsection{Test 2: the full 3D field}
\label{sec:sign_3d}

The symmetries of Sect.~\ref{sec:degeneracy} hold for any field, axisymmetric
or not, so sampling the full non-axisymmetric field cannot by itself supply the
sense. We nevertheless repeated the reconstruction on full 3D cubes
(no azimuthal average) on the funnel lattice, for three snapshots each of the
MAD and SANE $a_*=+0.94$ models. The handedness-blind geometry is
$\langle|\cos|\rangle_G=0.87$ (MAD) and $0.93$ (SANE), close to the mean-field
values ($0.89$ and $0.92$): the non-axisymmetric fluctuations are not
constrained by two one-dimensional profiles, and the reconstruction recovers
the mean structure. The signed alignment is $-0.004$ and $0.10$, again set by
the initial guess. All snapshots have more than enough jet cells; none is
excluded, with or without averaging.

\subsection{Test 3: a spin-based prior}
\label{sec:sign_spin}

\begin{figure}
\centering
\includegraphics[width=\columnwidth]{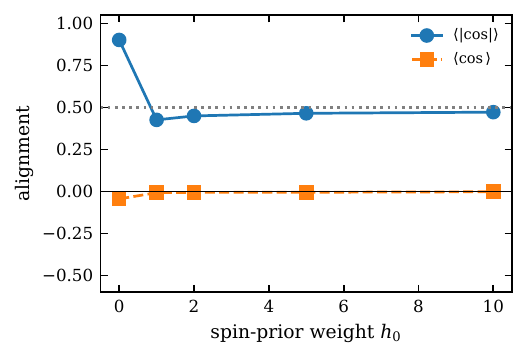}
\caption{The spin prior of Eq.~(\ref{eq:hspin}) for a retrograde model
(MAD, $a_*=-0.94$, $t=25145\,GM\,c^{-3}$): unsigned alignment $\langle|\cos|\rangle$ (circles) and
signed alignment $\langle\cos\rangle$ (squares) against the prior weight.
The prior destroys the orientation rather than fixing the sense.}
\label{fig:spincollapse}
\end{figure}

The third test assumes that the poloidal sense is tied to the black-hole
spin. We implement it as the frame-dragging term of the prior,
$\mathcal{H}_{\rm frame}=-\sum h\,S_z$ (Eq.~\ref{eq:ham} and
Appendix~\ref{app:prior}), with
\begin{equation}
h(\mathbf{x})=h_0\,\mathrm{sign}(a_*)\,\mathrm{sign}(z),
\label{eq:hspin}
\end{equation}
which rewards a poloidal field pointing away from the black hole for
prograde spin and towards it for retrograde spin; $h_0$ sets its weight. We
apply it to the retrograde model MAD $a_*=-0.94$, for which the default
reconstruction has $\langle|\cos|\rangle=0.90$. As soon as the prior is
switched on ($h_0\ge1$), the geometry collapses to
$\langle|\cos|\rangle=0.43$--$0.47$ ($\langle|\cos|\rangle_G=0.48$), while
$\langle\cos\rangle$ stays at zero (Fig.~\ref{fig:spincollapse}). The same
holds for three further snapshots ($0.43$). A wrong sense prior breaks the
orientation rather than flipping the sense.

There are two reasons for this failure, and neither is resolution. First,
Eq.~(\ref{eq:hspin}) has the parity of a split monopole, $B_z\propto\mathrm{sign}(z)$,
while the simulated jets are dipole-like (Sect.~\ref{sec:sign_spatial}), so for any
sign of $h_0$ it is wrong in one hemisphere. Second, and more importantly,
the Blandford--Znajek mechanism does not tie the sign of the poloidal field
to the spin. The extracted power is even in the field,
$P_{\rm BZ}\propto\Phi_{\rm BH}^2\,\Omega_{\rm H}^2$, and the black hole drives
the same outflow whichever way the flux threads it
\citep{bz77,tchekhovskoy2015}. The poloidal sense is inherited from the
polarity of the accreted flux, i.e.\ from the initial field of the
simulation. What the rotation does fix is the sign of the toroidal field
\emph{relative} to the poloidal field: field lines rotating with angular
velocity $\Omega_{\rm F}$ are swept back so that, in the outflow,
$B_\phi\simeq-(\Omega_{\rm F}R/c)\,B_{\rm p}$, with $B_{\rm p}$ measured along the
direction of the flow. The handedness of the helix,
which the mirror $\mathcal{T}$ reverses, is therefore set by the sense of
rotation, whereas the overall sign, which $-1$ reverses, is not. A spin prior
can at most break $\mathcal{T}$, and only if it is written as a handedness
prior (even in $\mathbf{B}$). $\mathcal{H}_{\rm frame}$ is odd in
$\mathbf{B}$ and so it is the wrong kind of term.

The prior acts on all cells, but the data term and the scoring exclude
$|z|\le3\,r_{\rm g}$. This region contains the ergosphere and the inner disc,
where the field is anchored and where the Blandford--Znajek and
Blandford--Payne mechanisms operate. The mask therefore removes the region
where the coupling between the field and the rotation of the hole or disc is
established. It does not remove the information on the sense: the poloidal
flux is frozen along field lines, and its sign in the jet is the sign with
which it threads the horizon. In any case the horizon radius
($r_+=1.35$--$2\,r_{\rm g}$) spans only two to three cells of the $96^3$ cube,
so the anchoring region could not be reconstructed even without the mask.
The information that the rotation imprints along the whole jet is the
handedness, and Sect.~\ref{sec:sign_route} uses it.

\subsection{The spin--sense relation in the simulations}
\label{sec:sign_diag}

\begin{figure}
\centering
\includegraphics[width=\columnwidth]{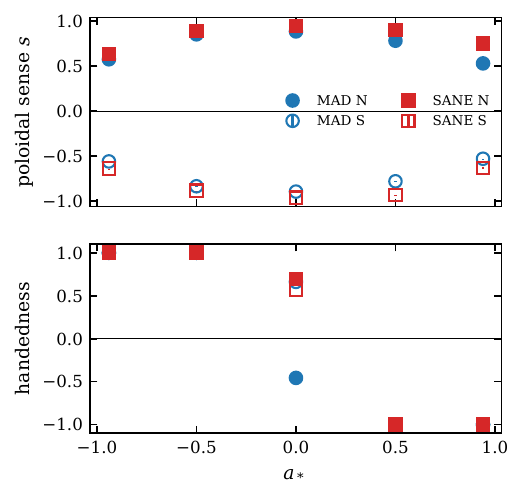}
\caption{Poloidal sense $s$ in the northern (filled) and southern (open)
jet, measured in the simulated fields, against black-hole spin, for MAD
(circles) and SANE (squares) models (\emph{upper panel}), and handedness
$\langle\mathrm{sign}(z\,B_\phi B_z)\rangle$ in the two jets (\emph{lower panel}).
The poloidal field points the same way in both jets in every model; the
handedness follows the sign of the spin.}
\label{fig:spinsense}
\end{figure}

To see what sets the poloidal sense in the simulations, we measure it
directly in the simulated fields, separately in the two jets:
$s_{\rm N,S}=\langle\mathrm{sign}(z)\,B_z/|\mathbf{B}|\rangle$, weighted by
$|\mathbf{B}|^2$ over the $\sigma>1$ region with $|z|>3\,r_{\rm g}$ ($s=+1$:
field pointing away from the black hole). We also measure the handedness
$\langle\mathrm{sign}(z\,B_\phi B_z)\rangle$ (Fig.~\ref{fig:spinsense}).
The result is the same in all ten models (Table~\ref{tab:fulldata},
Fig.~\ref{fig:spinsense}). In the northern jet $s_{\rm N}=+0.53$ to $+0.95$,
and in the southern jet $s_{\rm S}=-0.53$ to $-0.96$. In every model, MAD and
SANE, at every spin, $B_z$ points towards $+z$ in both jets: the configuration
is dipole-like and its polarity is the same throughout the library. This is
the polarity of the initial magnetic field of the simulations, which is
common to all models, carried by the accreted flux. An earlier version of this
diagnostic averaged $\mathrm{sign}(z)B_z$ over both jets. For a dipole-like
field that average measures only the imbalance between the hemispheres, which
is why it appeared to vanish in the SANE models. Measured per jet, the SANE
fields are as coherent as the MAD ones.

The magnitude of $s$ is not monotonic in spin, but it is nearly symmetric in
$|a_*|$. It is largest at $a_*=0$ ($0.89$ for MAD, $0.95$ for SANE) and
smallest at $|a_*|=0.94$ ($0.53$--$0.58$ for MAD, $0.63$--$0.75$ for SANE).
Since $s$ measures $B_z/|\mathbf{B}|$, this reflects the winding of the field:
the faster the hole rotates, the stronger the toroidal field generated by the
rotating field lines and the smaller the axial fraction. Its \emph{sign} does
not change with spin in either state, as expected if the spin does not set the
poloidal sense.

The handedness, by contrast, follows the spin exactly. In all eight models
with $a_*\neq0$, both jets have $\langle\mathrm{sign}(z\,B_\phi B_z)\rangle=-1$
for prograde spin and $+1$ for retrograde spin. Both states behave the same
way, including the retrograde models, in which the hole rotates against the
disc. This is the sign expected from $B_\phi\simeq-(\Omega_{\rm F}R/c)B_{\rm p}$
with $\Omega_{\rm F}$ following the rotation of the hole, and it shows that the
jet field lines are wound by the black hole rather than by the disc. At
$a_*=0$ the handedness is not definite: it is $-0.46$ and $+0.66$ in the two
MAD jets and $+0.70$ and $+0.58$ in the SANE ones.

\subsection{A route that works: rotation measure plus handedness}
\label{sec:sign_route}

Propositions~2 and 3 suggest a constructive route. The RM fixes the
toroidal sign. If the handedness is known --- from the sense of rotation of
the field lines, i.e.\ from the direction of the black-hole spin or of the
disc rotation --- the toroidal sign fixes the poloidal sign as well. We
add to the cost a handedness term,
$-w_{\rm h}\langle\mathrm{sign}(z)\,S_\phi S_z\rangle_{|\mathbf{B}|^2}$,
which is even under $-1$ and odd under $\mathcal{T}$, together with the RM
term. The cost then distinguishes the four images. On the synthetic jet, the
run started from the true image reaches $\langle\cos\rangle=0.996$, and the
runs started from the other three end at a cost higher by $13$--$25\%$
(Table~\ref{tab:symtest}). Gradient descent does not always cross the barrier
between images within 400 iterations, so those runs end in mixed states
($\langle\cos\rangle$ between $-0.35$ and $0.59$). Because the cost is no longer
degenerate, the remedy is a multi-start: running the four initial guesses and
keeping the lowest-cost solution selects the truth, without any knowledge of
it.

On the GRMHD data the handedness sign is known from
Sect.~\ref{sec:sign_diag}: it is $-\mathrm{sign}(a_*)$. A single run from the
default initial guess, whose handedness is wrong for the prograde models, does
not cross the barrier. For three snapshots of each $a_*=+0.94$ model it ends
at $\langle\cos\rangle=-0.015$ (MAD) and $0.03$ (SANE), and in MAD the geometry
is degraded ($\langle|\cos|\rangle_G=0.40$). The simulated field is dipole-like
with a toroidal field that reverses between the jets, so none of the four
images of the default guess is close to it. We therefore run the
multi-start over the sixteen sign patterns of $(B_z,B_\phi)$ in the two jets,
with $n_e\propto\rho$. For the representative snapshot (MAD, $a_*=+0.94$) the
lowest final cost is reached from the guess that has the pattern of the
simulated field: dipole-like, with $B_\phi\propto-\mathrm{sign}(z)$. The result has
$\langle\cos\rangle=+0.84$, so the poloidal sense is recovered on GRMHD data. The
runner-up is its reverse, whose cost is higher by only $0.7\%$, but whose
rotation-measure residual is $3.6$ times larger. The other fourteen patterns
end at costs higher by $1.7$--$11\%$. The margin between the truth and its
reverse is small because the handedness term, identical for both, dominates
the total cost. The discrimination comes from the RM alone, and it is clear in
the RM residual. A heavier RM weight, or a multi-frequency RM, should widen
it. What makes this work is the combination of an observable for the toroidal
sign (the RM) with physical knowledge of the sense of rotation. A finer grid
plays no role.

\subsection{Test 4: resolved multi-frequency polarimetry with internal Faraday rotation}
\label{sec:sign_ftomo}

The rotation-measure term of Sect.~\ref{sec:sign_rm} treats Faraday rotation
as an external screen: only the total Faraday depth
$\Phi(x,v)=\int n_e B_\parallel\,ds$ along each line of sight enters. In a
jet, emission and rotation occur in the same plasma. The polarization
observed at wavelength $\lambda$ is then
\begin{align}
P(x,v;\lambda^2)&=-\int \varepsilon\,
\frac{(B_x+iB_v)^2}{|\mathbf{B}_\perp|^2}\,
e^{2i\lambda^2\phi(s)}\,ds \nonumber\\
&=\int F(\phi)\,e^{2i\lambda^2\phi}\,d\phi ,
\label{eq:ptomo}
\end{align}
with $P=Q+iU$, where $s$ runs along the line of sight towards the observer,
$\phi(s)=\int_s^{\infty}n_eB_\parallel\,ds'$ is the Faraday depth between the
emitting element and the observer, $(x,v)$ are sky coordinates and $F$ is
the Faraday dispersion function of the line of sight \citep{burn1966}.
Resolved maps of $P$ at several wavelengths therefore carry information
on how emission and rotation are ordered in depth, which the RM map
discards.

\paragraph{Proposition 4 (internal Faraday rotation).} \emph{With the data
$P(x,v;\lambda^2)$ of Eq.~(\ref{eq:ptomo}) at the observed wavelengths, the
elements of $G$ act on the Faraday dispersion function of every line of
sight as}
\begin{equation}
\begin{aligned}
-1&:\ F(\phi)\to F(-\phi),\qquad
\mathcal{T}:\ F(\phi)\to F(\phi+\Phi),\\
-\mathcal{T}&:\ F(\phi)\to F(\Phi-\phi).
\end{aligned}
\label{eq:prop4}
\end{equation}
\emph{Equivalently, $P'(\lambda^2)=P(-\lambda^2)$,
$e^{-2i\lambda^2\Phi}P(\lambda^2)$ and $e^{2i\lambda^2\Phi}P(-\lambda^2)$,
respectively. The data are invariant under $-\mathcal{T}$ only if $F$ is
symmetric about $\Phi/2$ on every line of sight.}

\paragraph{Proof.} Under $-1$ the emissivity and $(B_x+iB_v)^2$ are
unchanged and $\phi\to-\phi$. Under $\mathcal{T}$ the emitting element at
depth $s$ moves to $-s$ with the same $(B_x+iB_v)^2$, and $B_\parallel$
changes sign, so its Faraday depth becomes
$-\int_{-\infty}^{s}n_eB_\parallel\,ds'=\phi(s)-\Phi$. Composing the two gives
$\Phi-\phi(s)$ for $-\mathcal{T}$. $\square$

\smallskip
The degeneracy that survives the RM, $-\mathcal{T}$, thus maps the Faraday
dispersion function to its reflection about $\Phi/2$, i.e.\ it exchanges the
near and the far side of the jet. It is exact only for configurations that
are symmetric in depth, such as a uniform slab \citep{burn1966}. For a
jet seen side-on, the radial (expansion) component of the field and any
departure from axisymmetry make the near and far sides differ, and the
degeneracy is lifted. At $\lambda=0$ all three relations reduce to
$P'=P$, which recovers Propositions~1 and~2. We verified
Eq.~(\ref{eq:prop4}) numerically: the relations hold to $10^{-13}$, while
$P$ itself changes by $O(1)$ under $-\mathcal{T}$.

To test whether the effect is usable, we extended the forward model to
resolved maps of $I$ and $P$ at four wavelengths, with internal Faraday
rotation, at arbitrary inclination (Appendix~\ref{app:rt}). The field is
resampled on an observer-aligned grid, and Eq.~(\ref{eq:ptomo}) is evaluated
along every line of sight. We used the synthetic jet of Sect.~\ref{sec:synth}
($\mu=1.2$, $64^3$ cells) with $n_e\propto|\mathbf{B}|$, and wavelengths chosen
so that the largest total rotation is $2\,$rad. At inclination $i\neq90^\circ$
the mirror through the sky plane is no longer $y\to-y$ but the reflection
$\mathcal{T}_k$ along the line of sight; Proposition~4 holds for it unchanged.
For each image of the true field ($-1$, $\mathcal{T}_k$, $-\mathcal{T}_k$) we
started the reconstruction from that image (with a 10\% random perturbation)
and let it converge, with and without Faraday rotation.
Table~\ref{tab:ftomo} gives the results.

Without Faraday rotation, all four starts converge to the same cost to
within the scatter of the optimizer ($5\%$ at $i=90^\circ$, $9\%$ at
$i=17^\circ$). This is the degeneracy of Propositions~1 and 2. With internal Faraday rotation, at $i=90^\circ$
every image ends with a cost $1.5$--$1.8$ times that of the true field, so a
multi-start minimization selects the truth. The same holds when the
reconstruction starts from the four sign patterns of the default guess
instead of the exact images (truth: $0.0074$; images: $0.016$--$0.020$;
control without rotation: $0.0072$--$0.0073$ for all four). At $i=17^\circ$ the
images $-1$ and $\mathcal{T}_k$ end $30\%$ above the truth, well outside the
scatter, so the field sense and the handedness are both fixed. The hardest case
is $-\mathcal{T}_k$, which reverses the poloidal field and the direction of
the jet in depth: its data distance from the truth is four times smaller
than for the other images, and its cost exceeds that of the truth by only
$9\%$, the same as the optimizer scatter at this inclination. In this test
$-\mathcal{T}_k$ is therefore not reliably excluded at $i=17^\circ$. For a
jet seen close to the axis the line of sight runs along the jet, the
distribution of Faraday depth is close to symmetric about $\Phi/2$, and
Proposition~4 then gives only a weak constraint. In practice the
approaching and receding jets are distinguished by Doppler boosting,
which is not part of the flat-space forward model used here; this breaks
$\mathcal{T}_k$ and $-\mathcal{T}_k$ independently of Faraday rotation
(Sect.~\ref{sec:disc_incl}).

\begin{table}
\caption{Resolved multi-frequency polarimetry with internal Faraday
rotation, synthetic jet. Each row starts from an exact image of the true
field. $D$: data distance of the image from the truth (summed normalized
residuals of $I$ and of $P$ at the four wavelengths); $\mathcal{L}$: final
cost with Faraday rotation ($\mathcal{L}_0$: without, $\lambda=0$ only).}
\label{tab:ftomo}
\centering
\begin{tabular}{l l c c c}
\hline\hline
$i$ & start & $D$ & $\mathcal{L}$ & $\mathcal{L}_0$ \\
\hline
$90^\circ$ & $\mathbf{B}$ (truth) & 0 & 0.0064 & 0.0064 \\
 & $-\mathbf{B}$ & 1.16 & 0.0119 & 0.0061 \\
 & $\mathcal{T}_k\mathbf{B}$ & 1.58 & 0.0098 & 0.0061 \\
 & $-\mathcal{T}_k\mathbf{B}$ & 1.31 & 0.0098 & 0.0064 \\
$17^\circ$ & $\mathbf{B}$ (truth) & 0 & 0.0063 & 0.0061 \\
 & $-\mathbf{B}$ & 0.58 & 0.0083 & 0.0057 \\
 & $\mathcal{T}_k\mathbf{B}$ & 0.57 & 0.0081 & 0.0063 \\
 & $-\mathcal{T}_k\mathbf{B}$ & 0.15 & 0.0069 & 0.0061 \\
\hline
\end{tabular}
\end{table}

\subsection{Test 5: ray-traced GRMHD images and synthetic ngEHT data}
\label{sec:sign_grrt}

\begin{figure}
\centering
\includegraphics[width=\columnwidth]{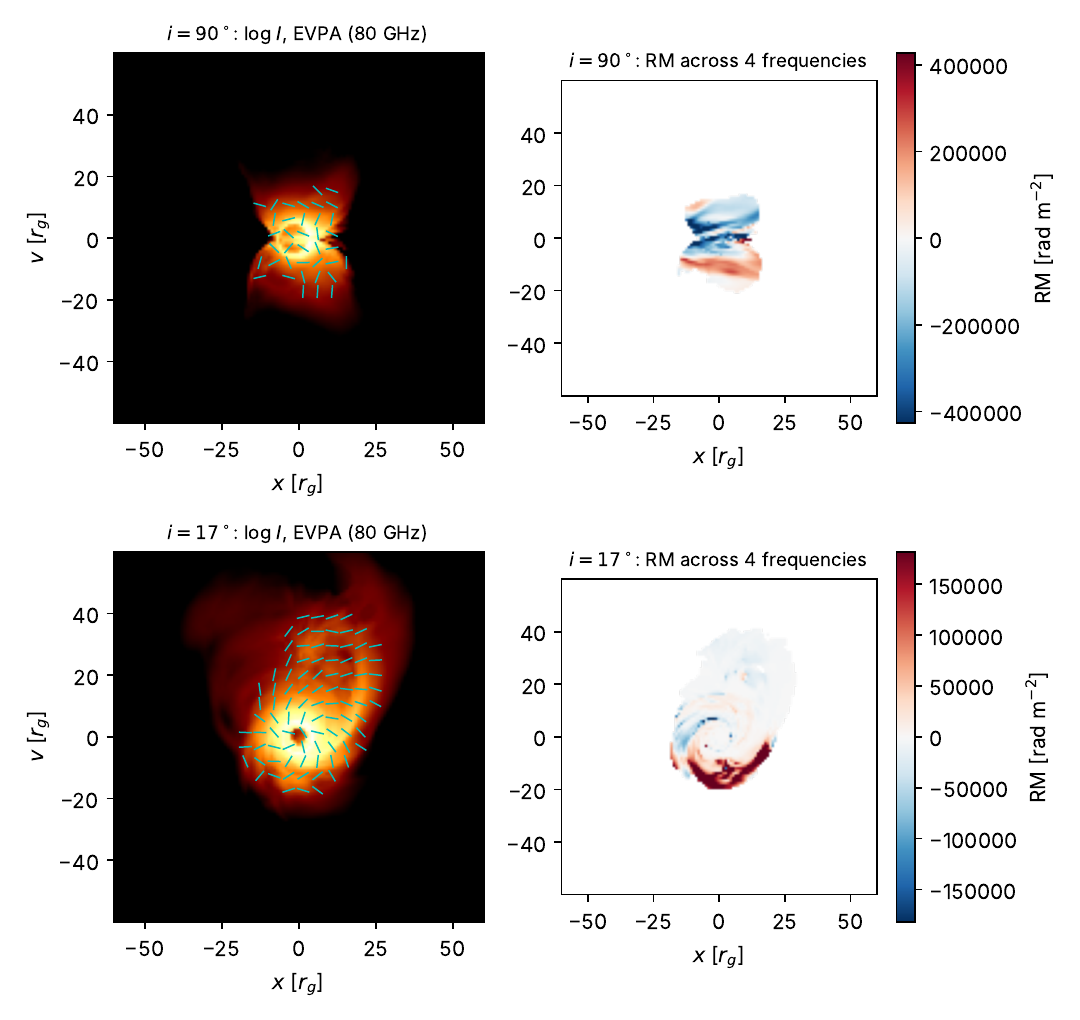}
\caption{Ray-traced 3\,mm images of the MAD$_1$ snapshot ($a=+0.94$)
(\texttt{ipole}, M87 mass and distance). Left: total intensity (logarithmic)
and polarization angle at 80\,GHz. Right: rotation measure fitted across the
four sub-bands (80--98\,GHz). Top: $i=90^\circ$; bottom: $i=17^\circ$.}
\label{fig:grrt}
\end{figure}

Test~4 uses the same forward model to generate and to invert the data. As a
test independent of that model, we inverted polarized images of GRMHD
snapshots computed with full radiative transfer. We used three snapshots
of the $a=+0.94$ runs of Sect.~\ref{sec:library}: the MAD snapshot at
$t=25\,215\,GM/c^3$ used so far (MAD$_1$), an earlier snapshot of the same
run at $t=25\,145\,GM/c^3$ (MAD$_2$), and the SANE snapshot at
$t=25\,145\,GM/c^3$. We
rendered them with \texttt{ipole} \citep{ipole2018}, with thermal synchrotron
emission, absorption, Faraday rotation and conversion along Kerr geodesics,
in four sub-bands at 3\,mm (80--98\,GHz), and MAD$_1$ also at 1.3\,mm
(214--262\,GHz), at $i=90^\circ$ and $i=17^\circ$ (Appendix~\ref{app:rt}).
In these images the
emission follows the plasma, not $|\mathbf{B}|^2$. The inversion therefore
fits a free emissivity together with the field, with $n_e$ proportional to
the emissivity and three global scales (flux, fractional polarization and
Faraday depth). The model knows nothing of the electron temperature, of the
emission physics or of the geodesics. We reconstructed the field on a
$64^3$ cube of half-size $30\,r_g$, masked the central $5\,r_g$ and, side-on,
the equatorial strip $|v|<3\,r_g$, and compared the result with the true
field of the snapshot. Figure~\ref{fig:grrt} shows the 3\,mm images of
MAD$_1$.

The amount of Faraday rotation within a band decides the outcome. As a
measure we use the flux-weighted rotation of the polarization angle across
the four sub-bands, $\langle|{\rm RM}|\,\Delta\lambda^2\rangle$. For MAD$_1$
it is $0.14$\,rad at 1.3\,mm and $0.92$\,rad at 3\,mm for $i=90^\circ$, and
$0.01$ and $0.08$\,rad for $i=17^\circ$, where the line of sight crosses the
tenuous funnel. At 3\,mm it is $0.77$ and $0.10$\,rad for MAD$_2$, and $1.20$
and $0.68$\,rad for the SANE snapshot. Table~\ref{tab:grrt} lists the final
costs when the reconstruction starts from the true field and from its images
under $G$, with and without Faraday rotation in the model.

For MAD$_1$ at 1.3\,mm the true field and its mirror image end at the same
cost ($0.0184$) at $i=90^\circ$. The rotation in the band is too small to be
used. At 3\,mm, side-on, the control without Faraday rotation gives again
identical costs for the true field and its mirror image ($0.3753$): the
degeneracy of Propositions~1 and 2 holds in the ray-traced data. With
internal Faraday rotation the mirror images end with a polarization residual
$2.5$--$3.7$ times that of the truth and a total cost $22$--$33\%$ higher.
In this snapshot the near--far ordering of Proposition~4 is thus measurable in
images computed with full radiative transfer, and it fixes the handedness.
The sense is not fixed: starting from $-\mathbf{B}$ the reconstruction
converges to a field with $\langle\cos\rangle=-0.78$ and the same cost as the
truth. All four solutions reproduce the sign of the RM on more than $95\%$
of the flux, because with a free emissivity the model can reverse
$B_\parallel$ only where the rotation occurs.

The other two snapshots show that this result is not general. In both, the
control without Faraday rotation again gives equal costs for the true field
and its mirror image, so that Faraday rotation is the only source of
separation, as Proposition~4 states. But the separation is small and does not
always favour the truth. For MAD$_2$, side-on, the four images end within
$18\%$ of each other, and the lowest cost ($0.0213$) is reached from
$-\mathcal{T}_k\mathbf{B}$, which has the wrong handedness, against $0.0219$
from the truth. For the SANE snapshot the fit is poorer (cost $0.144$ from
the truth, dominated by the polarization residual), and the mirror image ends
only $4\%$ higher. At $i=17^\circ$ the polarization residuals of the true
field and of its mirror image are equal without Faraday rotation, and close
with it for both MAD snapshots, where the rotation is small; the mirror image
ends with a higher cost mainly because $\mathcal{T}_k$ tilts the jet axis,
which the prior penalizes. In the SANE snapshot, with $0.68$\,rad of rotation,
Faraday rotation raises the polarization residual of the mirror image by
$21\%$.

A blind reconstruction does not reach these solutions. We used the four
sign patterns of the default guess, first without Faraday rotation (which
leaves only the degeneracy $G$) and then with it, starting from the four
images of the best field. For MAD$_1$ at 3\,mm, side-on, the best blind
solution ends at cost $0.060$ ($\langle|\cos|\rangle=0.60$). Continuing the
minimization four times longer from that field and from its mirror image
brings the costs to $0.0209$ and $0.0220$, close to that of the truth. The
lower of the two has the correct handedness ($\langle|\cos|\rangle=0.63$,
against $0.40$ for its mirror image), but by a margin of only $5\%$, and the
field has no definite sense ($\langle\cos\rangle=-0.01$). At $i=17^\circ$ the
best blind solution has a cost below that of the truth ($0.012$ against
$0.016$) with $\langle|\cos|\rangle=0.36$. We also started from the
reconstruction from $W$ and $p$ alone, made with the pipeline of
Sect.~\ref{sec:library}, and from its
three images under $G$. For MAD$_1$ that reconstruction is close to the
mirror image of the truth ($\langle|\cos|\rangle=0.23$, and $0.89$ against
the mirror image). After 1600 iterations the lowest cost ($0.029$) is reached
from its reversed field, with the wrong handedness, and the start with the
correct handedness ends at $0.037$, both well above the cost of the truth.
For MAD$_2$ and the SANE snapshot the blind and the $W,p$ starts end at
$1.4$--$14$ times the cost of the truth, again without a preference for the
correct handedness. With a free emissivity, four sub-bands and this
resolution, the cost therefore has distinct minima of nearly equal depth, and
the differences between the images of the truth are of the same order as the
differences between minima. The symmetry information enters the data only
through Faraday rotation, but in these ray-traced images it is not a robust
discriminant. Using it requires a better constrained emissivity (for instance
from a physical model of the electrons), a wider range of $\lambda^2$, or
several snapshots or time-averaged images.

Finally, we observed the 3\,mm side-on images of MAD$_1$ with the ngEHT
reference array using \texttt{ngehtsim} (Appendix~\ref{app:rt}). The common
restoring beam is $74\,\mu$as, or $20\,r_g$ for M87, so the emitting region
spans only about two beams and the beam depolarizes the source from
$m\simeq0.33$ to $0.03$. The rotation across the band survives
($0.88$\,rad). As a best case we first inverted the true images convolved
with this beam, with the same beam in the model. Starting from the true
field, the fit is much poorer than at full resolution (cost $0.19$,
$\langle\cos\rangle=0.58$), but the separation persists: the mirror image
ends at cost $0.29$, with a polarization residual $1.85$ times larger, while
without Faraday rotation the two end at the same cost ($0.435$). We then
reconstructed the images from the synthetic visibilities with
\texttt{ehtim} \citep{chael2018}. Restored with the same beam, the
\texttt{ehtim} maps differ from the convolved true images by $2\%$ in $I$
and $4$--$5\%$ in $P$. This is enough to reverse the result: the mirror
image now ends below the truth ($0.232$ against $0.244$), while without
Faraday rotation the two still end at the same cost ($0.437$). At 3\,mm and
ngEHT resolution, the handedness information of a side-on horizon-scale jet
is therefore below the imaging errors. Using it would require polarimetric
images accurate to about one per cent, and the method is better suited to
sources, or to scales, that span many beams.

\begin{table}
\caption{Ray-traced snapshots ($a=+0.94$; MAD$_1$: $t=25\,215$, MAD$_2$
and SANE: $t=25\,145\,GM/c^3$): final cost $\mathcal{L}$, polarization
residual $\mathcal{L}_P$ and $\langle\cos\rangle$ against the true field,
for reconstructions started from the true field and from its images.
$\mathcal{L}_0$: same, with a model without Faraday rotation. The blind row
is the best of the multi-start reconstruction (MAD$_1$ at $90^\circ$: after
1600 iterations). ngEHT: MAD$_1$ images convolved with the common restoring
beam of the synthetic coverage (beam), or reconstructed with \texttt{ehtim}
and restored with the same beam.}
\label{tab:grrt}
\centering
\setlength{\tabcolsep}{4pt}
\begin{tabular}{l l c c c c}
\hline\hline
data & start & $\mathcal{L}$ & $\mathcal{L}_P$ & $\langle\cos\rangle$ & $\mathcal{L}_0$ \\
\hline
MAD$_1$ 1.3\,mm, $90^\circ$ & $\mathbf{B}$ & 0.0184 & 0.0008 & $+0.73$ & -- \\
 & $\mathcal{T}_k\mathbf{B}$ & 0.0184 & 0.0007 & $-0.40$ & -- \\
MAD$_1$ 3\,mm, $90^\circ$ & $\mathbf{B}$ & 0.0199 & 0.0026 & $+0.78$ & 0.3753 \\
 & $-\mathbf{B}$ & 0.0190 & 0.0020 & $-0.78$ & -- \\
 & $\mathcal{T}_k\mathbf{B}$ & 0.0243 & 0.0064 & $-0.11$ & 0.3753 \\
 & $-\mathcal{T}_k\mathbf{B}$ & 0.0252 & 0.0074 & $+0.16$ & -- \\
 & blind & 0.0209 & 0.0032 & $-0.01$ & -- \\
MAD$_1$ 3\,mm, $17^\circ$ & $\mathbf{B}$ & 0.0156 & 0.0013 & $+0.91$ & 0.0223 \\
 & $-\mathbf{B}$ & 0.0155 & 0.0014 & $-0.88$ & -- \\
 & $\mathcal{T}_k\mathbf{B}$ & 0.0265 & 0.0010 & $-0.67$ & 0.0336 \\
 & $-\mathcal{T}_k\mathbf{B}$ & 0.0187 & 0.0010 & $+0.70$ & -- \\
 & blind & 0.0121 & 0.0030 & $+0.02$ & -- \\
MAD$_2$ 3\,mm, $90^\circ$ & $\mathbf{B}$ & 0.0219 & 0.0018 & $+0.77$ & 0.3108 \\
 & $-\mathbf{B}$ & 0.0230 & 0.0029 & $-0.77$ & -- \\
 & $\mathcal{T}_k\mathbf{B}$ & 0.0252 & 0.0044 & $-0.14$ & 0.3110 \\
 & $-\mathcal{T}_k\mathbf{B}$ & 0.0213 & 0.0020 & $+0.18$ & -- \\
 & blind & 0.0716 & 0.0275 & $-0.06$ & -- \\
MAD$_2$ 3\,mm, $17^\circ$ & $\mathbf{B}$ & 0.0155 & 0.0010 & $+0.88$ & 0.0280 \\
 & $\mathcal{T}_k\mathbf{B}$ & 0.0543 & 0.0017 & $-0.61$ & 0.0653 \\
SANE 3\,mm, $90^\circ$ & $\mathbf{B}$ & 0.144 & 0.098 & $+0.63$ & 0.790 \\
 & $-\mathbf{B}$ & 0.262 & 0.216 & $-0.63$ & -- \\
 & $\mathcal{T}_k\mathbf{B}$ & 0.149 & 0.103 & $-0.15$ & 0.791 \\
 & $-\mathcal{T}_k\mathbf{B}$ & 0.162 & 0.117 & $+0.16$ & -- \\
 & blind & 0.304 & 0.281 & $+0.18$ & -- \\
SANE 3\,mm, $17^\circ$ & $\mathbf{B}$ & 0.115 & 0.064 & $+0.60$ & 0.269 \\
 & $\mathcal{T}_k\mathbf{B}$ & 0.154 & 0.077 & $-0.24$ & 0.296 \\
ngEHT beam & $\mathbf{B}$ & 0.190 & 0.119 & $+0.58$ & 0.435 \\
 & $\mathcal{T}_k\mathbf{B}$ & 0.292 & 0.220 & $-0.26$ & 0.436 \\
ngEHT \texttt{ehtim} & $\mathbf{B}$ & 0.244 & 0.175 & $+0.68$ & 0.437 \\
 & $\mathcal{T}_k\mathbf{B}$ & 0.232 & 0.157 & $-0.26$ & 0.437 \\
\hline
\end{tabular}
\end{table}

\subsection{Synthesis}
\label{sec:sign_synth}

Table~\ref{tab:tests} summarizes the tests. The failures are all explained by
the symmetries of the observables and by the physics of the simulated jets,
and none of them is due to how
the simulations were sampled. The sense of the poloidal field is not
recoverable from $W$ and $p$. For a side-on jet it is not recoverable from
the RM either, which only adds the toroidal sign. It becomes recoverable
once the handedness is fixed independently, or once the data resolve how
emission and Faraday rotation are ordered along the line of sight
(Proposition~4), or once the viewing angle brings $B_z$ onto the line of
sight (Sect.~\ref{sec:disc_incl}). The ray-traced images of Test~5 confirm
outside our own forward model that only Faraday rotation separates the
images of $G$, but they also show the limits of the second route: in three
GRMHD snapshots at 3\,mm, side-on, the lowest cost with the wrong
handedness lies between $3\%$ below and $22\%$ above that of the truth, and
at ngEHT resolution the separation is lost in the imaging errors.

\begin{table}
\caption{The routes tested to recover the field sense on the GRMHD data, and
what limits each.}
\label{tab:tests}
\centering
\begin{tabular}{p{0.28\columnwidth} p{0.60\columnwidth}}
\hline\hline
Route & Outcome and reason \\
\hline
RM term in loss & Fixes the toroidal sign, not the poloidal sign:
side-on, $B_z$ does not enter the RM (Proposition~3). Independent of
resolution. \\
Full 3D sampling & The symmetries hold for any field; the sense remains
that of the initial guess. \\
Spin prior ($\mathcal{H}_{\rm frame}$) & Spin does not set the poloidal sign;
the prior also has the wrong (monopole) parity for dipole-like jets. \\
RM + handedness & Makes the cost non-degenerate; a multi-start
minimization recovers the sense when the rotation sense of the field lines is
known ($\langle\cos\rangle=0.996$ synthetic, $0.84$ GRMHD). \\
Resolved multi-$\lambda$ polarimetry & Internal Faraday rotation breaks all
of $G$ (Proposition~4); the multi-start selects the truth on synthetic jets
at $i=90^\circ$. At $i=17^\circ$, $-\mathcal{T}_k$ is only weakly
disfavoured. \\
Ray-traced GRMHD (\texttt{ipole}) & Without Faraday rotation the images
of $G$ are exactly degenerate. Side-on at 3\,mm, Faraday rotation separates
them, but not robustly: the lowest cost with the wrong handedness lies
$22\%$ (MAD$_1$) and $4\%$ (SANE) above that of the truth, and $3\%$ below it
(MAD$_2$); the sense is not fixed. At 1.3\,mm and at $i=17^\circ$
(MAD) the rotation within the band is too small. Blind and $W,p$-informed
starts do not reach the truth. With \texttt{ehtim} images at ngEHT
resolution the mirror image is preferred. \\
\hline
\end{tabular}
\end{table}

\section{Discussion}
\label{sec:discussion}

\subsection{What H-MOG achieves}
\label{sec:disc_achieve}

H-MOG reconstructs the 3D orientation of the magnetic field from two
projected observables --- width and linear polarization --- with accuracy
$\langle|\cos|\rangle_G\simeq0.76$--$0.98$ across ten GRMHD models
spanning both accretion states and the full spin range. The reconstruction is
a free field on a lattice, not a low-dimensional parametric fit. No symmetry
is imposed beyond what the prior gently encourages, and no training set is
used. The Hamiltonian prior makes the ill-posed inverse problem tractable,
and the robustness tests of Sect.~\ref{sec:synth_noise} show that it does
so without overfitting. What is recovered is the geometry up to the four
images of Sect.~\ref{sec:degeneracy}. The image is chosen by the initial
guess, and any statement about the sense or the handedness of a reconstructed
field must say so.

\subsection{The nature of the limit}
\label{sec:disc_limit}

The limit on the sense is a property of the observables, not of the
optimizer. $W$ and $p$ are invariant under field reversal and under
reflection through the plane of the sky (Propositions~1 and 2), so no
procedure using only these observables can distinguish the four images. This
degeneracy is not fundamental in any broader sense. It holds
for parity-even, line-of-sight-integrated observables of a side-on jet, and
it is lifted step by step by adding observables or physical knowledge
(Sect.~\ref{sec:disc_routes}). The tests of Sect.~\ref{sec:sign} show that
the routes we tried are limited by these symmetries and not by the sampling of
the simulations. They also show something about the simulations: the
poloidal sense in these jets is not set by the spin
(Sect.~\ref{sec:sign_diag}). A method that recovered the sense by building in a
spin assumption would have misrepresented the physics of these flows.

\subsection{Dependence on the viewing angle}
\label{sec:disc_incl}

All results above are for a viewing angle $i=90^\circ$. At other
inclinations two things change. First, the mirror $\mathcal{T}$ still leaves
$W$ and $p$ invariant, but it now maps a jet at inclination $i$ to one at
$180^\circ-i$, i.e.\ it exchanges the approaching and the receding jet. Once
relativistic beaming is included in the forward model, the brightness ratio of
jet and counter-jet distinguishes the two, and $\mathcal{T}$ is broken by the
geometry itself. The handedness becomes measurable in principle. Second,
the axial field acquires a line-of-sight component $B_z\cos i$, so that
the RM becomes sensitive to the poloidal sense. Proposition~3 then no longer
protects $-\mathcal{T}$, and the RM can fix the sense outright. The side-on case
treated here is therefore the most degenerate one. For M87 ($i\simeq17^\circ$)
the axial field enters the line-of-sight component with weight
$\cos i\simeq0.96$, so Faraday rotation is directly sensitive to the poloidal
sense, unlike for a side-on jet. Small
inclinations also bring their own difficulties: the projected width is
foreshortened and mixes heights along the jet, the polarization is dominated
by the toroidal field, and beaming weights the near side of the jet. The resolved forward model of
Tests~4 and 5 already works at arbitrary $i$, without beaming. In the
ray-traced MAD images at $i=17^\circ$ the line of sight runs along the
tenuous funnel, the rotation within a band is small, and the mirror image is
disfavoured mainly by the prior on the jet axis (Sect.~\ref{sec:sign_grrt}). Adding
relativistic beaming to the forward model is the natural next step, and the
symmetry analysis above tells in advance which degeneracies to expect.

\subsection{Comparison with existing approaches}
\label{sec:disc_compare}

H-MOG sits among several families of methods for inferring jet magnetic
structure, and it is useful to state precisely how it differs from each.

\paragraph{Parametric jet models.} A long tradition fits the observed
emission with an assumed field geometry --- a helix of prescribed pitch, a
self-similar magnetocentrifugal wind, a force-free configuration anchored
to the horizon \citep{blandford_payne82,vlahakis2004,nakamura2018}. These
models can produce a field with a definite sense, but only because the
sense is built into the assumed configuration; they constrain a handful of
parameters, not a free field, and they cannot reveal structure the
parametrization does not contain. H-MOG instead reconstructs the field as
a free function on a lattice.

\paragraph{Rotation-measure mapping.} Faraday rotation is linear in the
line-of-sight field, and a transverse gradient of the rotation measure
across a jet has long been used as evidence for a helical, and hence
signed, toroidal component
\citep{zavala_taylor2004,broderick_mckinney2010,homan2009}. Proposition~3
makes the scope of this inference precise: for a jet seen side-on the RM
gradient fixes the sign of the toroidal field, and the poloidal sign requires
either a significant line-of-sight component of $B_z$ or the rotation sense of
the field lines.

\paragraph{Forward-modelling and GRMHD fitting.} The EHT
programme compares polarized images directly against large
libraries of GRMHD snapshots, selecting the models consistent with the
data \citep{eht2021_viii,dhruv2025}. This is powerful for constraining
bulk parameters --- magnetic state, spin, inclination --- but it infers the
field by selecting a simulation rather than reconstructing it, so the
recovered field is only ever as rich as the library, and the sense it
reports is the sense of the chosen simulation. H-MOG reconstructs a field
not present in any library, and states which properties of it no parity-even
comparison can fix.

\paragraph{Field extrapolation.} In the solar corona, 3D fields are
reconstructed by extrapolating a measured boundary field under a force-free
assumption. The philosophy is similar --- physical regularization selects
among many fields consistent with limited data --- but the inputs differ.
Coronal extrapolation starts from a measured vector magnetogram, while for
jets only projected, parity-even observables are available.

\subsection{Routes to break the degeneracy}
\label{sec:disc_routes}

Recovering the sense requires observables or knowledge that break the
symmetries of Sect.~\ref{sec:degeneracy}. Faraday rotation breaks $-1$ and
$\mathcal{T}$. Combined with the sense of rotation of the field lines, or with
a viewing angle that puts $B_z$ on the line of sight, it gives the full sense.
Faraday tomography, which uses the wavelength dependence of the
polarization angle to separate contributions along the line of sight, can
in addition locate the field in depth, given a model for the electron density
\citep{zavala_taylor2004}. Circular polarization (Stokes $V$) is natively
odd in the field, since it depends on the sense of the electron gyration,
and is therefore a direct sense diagnostic where it can be measured
\citep{eht2023m87circ}. Cross-correlations between the velocity field and
the magnetic field, through the frozen-in condition of ideal MHD, also
break the parity, since the induction term $\mathbf{v}\times\mathbf{B}$
changes sign with the field.

A complete test of these routes requires polarized radiative transfer, in
which the emission model fixes the electron density and Faraday rotation and
conversion are computed self-consistently. Test~5 is a first step: it inverts
\texttt{ipole} images of three GRMHD snapshots and synthetic ngEHT observations
of one of them. It shows that the degeneracy holds exactly in ray-traced
images without Faraday rotation, that the 1.3\,mm band is too narrow in
$\lambda^2$ to break it, and that at 3\,mm the internal Faraday rotation
separates the images of $G$ but not robustly: with a free emissivity the cost
has several minima of nearly equal depth, the true handedness is preferred in
two snapshots out of three, and the preference does not survive
\texttt{ehtim} imaging at ngEHT resolution. A physical emissivity model in
the inversion, wider bands and time-averaged images are the obvious
improvements. A systematic
study --- over the GRMHD library, inclinations, frequencies and
time-averaged images, with relativistic beaming in the forward model --- is
left for future work. The application to real VLBI data, where
signal-to-noise, finite beam size and the bright knots and recollimation
shocks outside the scope of the method all enter, will follow the same path.

\subsection{Observational prospects}
\label{sec:disc_prospects}

H-MOG inverts width and linear polarization --- exactly what VLBI and the
EHT now deliver on horizon scales \citep{eht2021_vii,eht2021_viii,kim2018}.
This makes geometric reconstruction applicable to current data, with the
proviso that the sense and the handedness of the reconstructed field are not
measured. Near-horizon circular polarization in M87* \citep{eht2023m87circ},
multi-frequency Faraday rotation, and the resolved polarimetry expected from
the ngEHT and from the Square Kilometre Array (SKA) are the most likely
sources of the missing information.

One corollary needs stating plainly. Polarization maps of M87*, Sgr~A* and
other jets will be read for magnetic structure: helical fields, collimation,
ordered winding. Our results suggest that the unsigned geometry so read can
be trusted.
Unless the analysis includes a parity-odd observable, however, the
\emph{sense} attached to that geometry --- the arrowhead on each field line ---
and, for a side-on view, the handedness of the helix will be choices made by
the reconstruction, not measured quantities. The degeneracy does not announce
itself. A method that fixes the sense silently, through initialization or
prior, returns a definite answer that looks like a measurement but is not.
We suggest that the sense be reported as undetermined until an observable
or an independent physical constraint fixes it.

\section{Conclusions}
\label{sec:conclusions}

We have presented H-MOG, a variational method that reconstructs the 3D
magnetic geometry of a relativistic jet from projected observables --- the
jet width and the linear polarization --- regularized by a Hamiltonian
prior and optimized with automatic differentiation. Our findings are the
following.

\begin{enumerate}

\item On synthetic jets and on a library of ten GRMHD models spanning both
accretion states (MAD and SANE) and five black-hole spins, H-MOG recovers the
field orientation, up to the mirror image, with accuracy
$\langle|\cos|\rangle_G\simeq0.76$--$0.98$ (synthetic jets: $0.94$--$1.00$), far
above the random value $0.5$, and robustly to observational noise up to the
level of the signal.

\item For a side-on jet, the width and polarization are invariant under field
reversal and under reflection through the plane of the sky
(Propositions~1 and 2). The poloidal sense and the handedness of the helix
are therefore not determined by these observables. The reconstruction
inherits both from its initial guess, which we verify directly.

\item A rotation-measure term fixes the toroidal sign but not the poloidal
sign (Proposition~3), at any lattice resolution. Full 3D sampling does not
change this. A spin prior fails because the Blandford--Znajek mechanism does
not tie the poloidal sign to the spin: in all ten simulations the poloidal
field points the same way in both jets at every spin, while the handedness of
the helix follows the sign of the spin.

\item The poloidal sense becomes recoverable when the RM is combined with the
rotation sense of the field lines, which fixes the handedness: a multi-start
minimization then selects the true image ($\langle\cos\rangle=0.996$ on a
synthetic jet, $0.84$ on a GRMHD snapshot), or, beyond the
side-on case, when the viewing angle brings the axial field onto the line of
sight. Circular polarization is a direct alternative.

\item Internal Faraday rotation breaks all of $G$ (Proposition~4). On
synthetic jets a multi-start minimization then selects the true image
side-on. On GRMHD images computed with full polarized radiative transfer,
the 3\,mm band fixes the handedness of a side-on jet but not the sense,
Without Faraday rotation the images of $G$ remain exactly degenerate, and at
1.3\,mm, or at $i=17^\circ$, the rotation within the band is too small to be
used. At 3\,mm, side-on, Faraday rotation separates the images, but in three
snapshots the true handedness is preferred by $22\%$, $4\%$ and $-3\%$ of the
cost, and in synthetic ngEHT data imaged with \texttt{ehtim} the preference is
lost. Internal Faraday rotation is thus the route that can break $G$, but with
a free emissivity it is not yet a robust discriminant.

\end{enumerate}

From the width and polarization of a jet, H-MOG recovers the unsigned 3D
geometry of its magnetic field without assuming a model. Which way that field
points, and which way it winds, are set by the symmetries of the observables
rather than by the algorithm. The symmetry analysis tells which additional
observable fixes each of them. Stating these limits
explicitly, rather than letting a modelling choice hide them, is what we
take to be the contribution of this work.


\begin{acknowledgements}

We thank the authors of the GRMHD simulation library of
\citet{dhruv2025} for making their data publicly available at
\texttt{http://thz.astro.illinois.edu/}. This work made use of
\texttt{ipole} \citep{ipole2018} for polarized ray tracing,
\texttt{ngehtsim} \citep{ngehtsim} and \texttt{ehtim} \citep{chael2018} for
synthetic interferometric data, \texttt{JAX} for automatic differentiation, and of
\texttt{NumPy}, \texttt{SciPy} and \texttt{Matplotlib}.
\end{acknowledgements}

\bibliographystyle{aa}

\bibliography{references}

\begin{appendix}

\section{Hamiltonian prior and initial guess: full expressions}
\label{app:prior}

The prior of Eq.~(\ref{eq:ham}) acts on the unit direction field
$\mathbf{S}=(S_x,S_y,S_z)$, with the field written as
$\mathbf{B}=A\,\mathbf{S}$. We give here the explicit lattice form of each
term. Sums run over all lattice cells, and the gradients are computed by
centred finite differences in lattice units.

\paragraph{Anti-kink (smoothness).} The dominant regularizing term
penalizes spatial variation of the field direction,
\begin{equation}
\mathcal{H}_{\rm kink} = \sum_{\rm cells}
\left( |\nabla S_x|^2 + |\nabla S_y|^2 + |\nabla S_z|^2 \right),
\end{equation}
and is weighted by $\lambda=1$ throughout.

\paragraph{External pressure (collimation).} The local opening angle of
the field with respect to the jet axis is $\theta=\arccos|S_z|$. The
external-pressure term draws it towards a height-dependent target,
\begin{equation}
\mathcal{H}_{\rm ext} = \sum_{\rm cells}
\bigl(\theta - \theta_{\rm t}(z)\bigr)^2,
\qquad
\theta_{\rm t}(z) = \theta_0
\left(\frac{|z|}{z_0}\right)^{-\frac{1}{2}(k-1)},
\end{equation}
with $\theta_0=8^\circ$, $z_0=50\,r_{\rm g}$ and $k=2$; this encodes a
collimating confinement of the kind inferred for the M87 jet
\citep{asada2011,nakamura2018}. It enters with weight $\alpha=0.05$.

\paragraph{Shear layer.} The shear term acts in a cylindrical shell,
\begin{equation}
\mathcal{H}_{\rm shear} = \sum_{\rm cells}
w(R)\,\bigl(|S_z| - \mu_{\rm t}\bigr)^2,
\end{equation}
\begin{equation}
w(R) = \exp\!\left[-\frac{(R/R_{\rm n}-R_0)^2}{2\,\sigma_R^2}\right],
\end{equation}
with target $\mu_{\rm t}=0.85$, $R_0=0.7$ and $\sigma_R=0.15$. The radius is
normalized by $R_{\rm n}=30\sqrt{2}\,r_{\rm g}$, the half-diagonal of the
$96^3$ cube, so the window is centred at $R\simeq30\,r_{\rm g}$. This lies
outside most of the funnel at all heights of the domain, which is why the shear weight has almost no effect (Sect.~\ref{sec:prior_sens}). It
enters with weight $\beta=0.3$.

\paragraph{Frame dragging.} A term linear in the axial field component,
\begin{equation}
\mathcal{H}_{\rm frame} = -\sum_{\rm cells} h\,S_z,
\end{equation}
is the only term of the prior that is odd in $\mathbf{S}$. We set $h=0$ in
every reconstruction except the spin-prior test of
Sect.~\ref{sec:sign_spin}, where $h$ is given by Eq.~(\ref{eq:hspin}).

\paragraph{Solenoidal penalty.} Separately from $\mathcal{H}$, the
divergence of the full field is penalized by
\begin{equation}
\mathcal{L}_{\rm div} = \frac{1}{N_{\rm cells}}\sum_{\rm cells}
\left(\partial_x B_x + \partial_y B_y + \partial_z B_z\right)^2 ,
\end{equation}
with derivatives in units of the $96^3$ lattice spacing, added to the total
cost with weight $w_{\rm div}=0.05$.

With $h=0$, every term above is invariant under $\mathbf{S}\to-\mathbf{S}$
and under the mirror $\mathcal{T}$ of Eq.~(\ref{eq:mirror}), the property
used in Propositions~1 and 2. The prior is divided by the number of cells
and enters the cost of Eq.~(\ref{eq:loss}) with weight
$\alpha_{\rm prior}=0.5$.

\paragraph{Initial guess.} The amplitude is initialized as a sheath,
$A_0\propto\exp\{-[(R/R_0(z)-1)/0.45]^2\}$ with $R_0(z)=0.8\,\ell\,(|z|/\ell)^{0.4}$,
and the direction as
$\mathbf{S}_0\propto(-0.3\,s_\phi\sin\phi,\;0.3\,s_\phi\cos\phi,\;0.9\,s_z\,\mathrm{sign}(z))+0.05\,\boldsymbol{\xi}$,
with $\boldsymbol{\xi}$ a unit Gaussian random field, $\ell=1$ lattice unit
for the synthetic jets and $\ell=3\,r_{\rm g}$ for the GRMHD data. The default
is $s_z=s_\phi=+1$. The four images of Sect.~\ref{sec:synth_sym} correspond to
$(s_z,s_\phi)=(\pm1,\pm1)$, and the dipole-like guess of
Sect.~\ref{sec:sign_spatial} replaces $\mathrm{sign}(z)$ with $1$. The sixteen
patterns of Sect.~\ref{sec:sign_route} combine these choices with a toroidal
component that is either of one sign in both jets, as above, or proportional
to $s_\phi\,\mathrm{sign}(z)$.

\section{GRMHD model parameters and native resolution}
\label{app:models}

Table~\ref{tab:models} summarizes the ten GRMHD models used in this work,
drawn from the library of \citet{dhruv2025}. All models share the native
resolution of $288\times128\times128$ in FMKS coordinates and were evolved to
$3\times10^4\,GM\,c^{-3}$. Table~\ref{tab:native} gives the native resolution
across the jet funnel.

\begin{table}[h]
\caption{The ten GRMHD models. The spin values are
$a_*=\pm15/16,\pm1/2,0$, quoted as $\pm0.94,\pm0.50,0$ in the text.}
\label{tab:models}
\centering
\begin{tabular}{l c c c}
\hline\hline
State & $a_*$ & Snapshots used & Time range [$GM\,c^{-3}$] \\
\hline
MAD  & $-15/16$ & 10 & 25100--25145 \\
MAD  & $-1/2$   & 10 & 25100--25145 \\
MAD  & $0$      & 10 & 25100--25145 \\
MAD  & $+1/2$   & 10 & 25100--25145 \\
MAD  & $+15/16$ & 10 & 25215--25260 \\
SANE & $-15/16$ & 10 & 25100--25145 \\
SANE & $-1/2$   & 10 & 25100--25145 \\
SANE & $0$      & 10 & 25100--25145 \\
SANE & $+1/2$   & 10 & 25100--25145 \\
SANE & $+15/16$ & 10 & 25100--25145 \\
\hline
\end{tabular}
\end{table}

\begin{table}[h]
\caption{Native resolution of the FMKS grid across the jet funnel
(MAD, $a_*=+0.94$): transverse size of the first polar cell,
$r\,\Delta\theta$, half-opening angle of the $\sigma>1$ funnel, and number of
native $\theta$ cells across it. The Cartesian spacing is $0.63\,r_{\rm g}$
($96^3$ cube) and $0.16\,r_{\rm g}$ (funnel lattice).}
\label{tab:native}
\centering
\begin{tabular}{c c c c}
\hline\hline
$r$ [$r_{\rm g}$] & $r\,\Delta\theta$ [$r_{\rm g}$] & $\theta_{\sigma=1}$ [rad] & $N_\theta$ \\
\hline
3  & $0.36$ & $1.57$ & $64$ \\
5  & $0.51$ & $1.43$ & $51$ \\
10 & $0.84$ & $1.15$ & $32$ \\
20 & $1.44$ & $0.83$ & $20$ \\
30 & $1.98$ & $0.69$ & $16$ \\
\hline
\end{tabular}
\end{table}

\section{Optimization convergence}
\label{app:conv}

\begin{figure}
\centering
\includegraphics[width=\columnwidth]{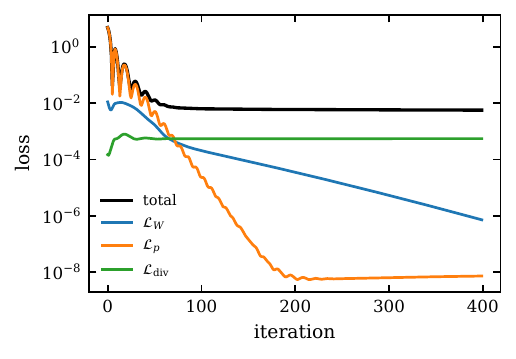}
\caption{Convergence of the H-MOG cost functional for the synthetic
reconstruction. The total loss (black) and its components --- width
$\mathcal{L}_W$, polarization $\mathcal{L}_p$, and divergence
$\mathcal{L}_{\rm div}$ --- are shown against iteration number on a
logarithmic scale.}
\label{fig:conv}
\end{figure}

\begin{figure}
\centering
\includegraphics[width=\columnwidth]{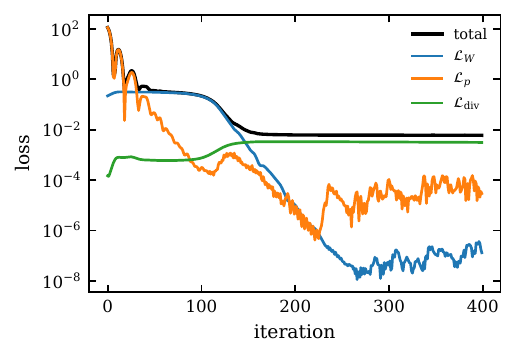}
\caption{As Fig.~\ref{fig:conv}, for the representative GRMHD reconstruction
(MAD, $a_*=+0.94$).}
\label{fig:conv_grmhd}
\end{figure}

The reconstruction is run for $400$ iterations of the Adam optimizer with
learning rate $0.05$. The total cost falls by three (synthetic) to four (GRMHD)
orders of magnitude, and the data residuals on $W$ and $p$ reach the
sub-percent level within the first $200$ iterations and are stationary
thereafter (Figs.~\ref{fig:conv} and \ref{fig:conv_grmhd}). Within the image
selected by the initial guess, the recovered geometry does not depend on the
details of the initialization: the four sign variants of the default guess
give the same $\langle|\cos|\rangle_G=0.89$ on the representative GRMHD
snapshot (Sect.~\ref{sec:sign_spatial}).

\section{Full GRMHD reconstruction results}
\label{app:data}

Table~\ref{tab:fulldata} lists the full per-model results of the H-MOG
reconstruction across the GRMHD library. For each of the ten models we report
the alignment scores of Sect.~\ref{sec:synth_setup}, the axis correlation
$\langle 2\cos^2\vartheta-1\rangle=\langle\cos2\vartheta\rangle$, and the sense
and handedness of the simulated field of Sect.~\ref{sec:sign_diag}. The axis
correlation is a second-order measure of alignment that, like
$\langle|\cos|\rangle$, is blind to the sense. It equals $1$ for perfect
alignment, $-1$ for a field everywhere perpendicular to the truth, and
$-1/3$ for random directions. It penalizes large misalignments more strongly
than $\langle|\cos|\rangle$ does. For the prograde models it is negative,
because a helix of the opposite handedness is nearly perpendicular to the
truth where the field is strongly wound.

\begin{table*}
\caption{Complete per-model statistics across the GRMHD library (mean-field cube,
$96^3$, default initial guess, means over $N_{\rm s}=10$ snapshots). Alignment
scores: $\langle|\cos|\rangle$, $\langle|\cos|\rangle_G$, axis correlation
$\langle\cos2\vartheta\rangle$ and signed alignment $\langle\cos\rangle$ (random
values $0.5$, $\simeq0.5$, $-1/3$ and $0$). Properties of the simulated field: poloidal
sense $s$ in the northern and southern jets and handedness
$\langle\mathrm{sign}(z B_\phi B_z)\rangle$ (N/S); see Sect.~\ref{sec:sign_diag}.
Snapshot scatters of $\langle|\cos|\rangle$ and $\langle|\cos|\rangle_G$ are given in
Table~\ref{tab:grmhd}.}
\label{tab:fulldata}
\centering
\begin{tabular}{l c c c c c c c c}
\hline\hline
State & $a_*$ & $\langle|\cos|\rangle$ & $\langle|\cos|\rangle_G$ & $\langle\cos2\vartheta\rangle$ & $\langle\cos\rangle$ & $s_{\rm N}$ & $s_{\rm S}$ & handedness N/S \\
\hline
MAD & $-0.94$ & $0.901$ & $0.901$ & $+0.63$ & $+0.03\pm0.04$ & $+0.58$ & $-0.56$ & $+1.00/+1.00$ \\
MAD & $-0.50$ & $0.889$ & $0.889$ & $+0.59$ & $-0.01\pm0.03$ & $+0.85$ & $-0.84$ & $+1.00/+1.00$ \\
MAD & $\phantom{+}0.00$ & $0.755$ & $0.761$ & $+0.18$ & $-0.02\pm0.01$ & $+0.89$ & $-0.90$ & $-0.46/+0.66$ \\
MAD & $+0.50$ & $0.439$ & $0.874$ & $-0.50$ & $+0.03\pm0.00$ & $+0.78$ & $-0.78$ & $-1.00/-1.00$ \\
MAD & $+0.94$ & $0.228$ & $0.890$ & $-0.86$ & $-0.01\pm0.00$ & $+0.53$ & $-0.53$ & $-1.00/-1.00$ \\
\hline
SANE & $-0.94$ & $0.921$ & $0.921$ & $+0.70$ & $-0.13\pm0.03$ & $+0.64$ & $-0.64$ & $+1.00/+1.00$ \\
SANE & $-0.50$ & $0.962$ & $0.962$ & $+0.85$ & $+0.29\pm0.01$ & $+0.89$ & $-0.89$ & $+1.00/+1.00$ \\
SANE & $\phantom{+}0.00$ & $0.926$ & $0.926$ & $+0.72$ & $+0.42\pm0.00$ & $+0.95$ & $-0.96$ & $+0.70/+0.58$ \\
SANE & $+0.50$ & $0.814$ & $0.979$ & $+0.36$ & $+0.07\pm0.01$ & $+0.90$ & $-0.93$ & $-1.00/-1.00$ \\
SANE & $+0.94$ & $0.318$ & $0.921$ & $-0.70$ & $+0.14\pm0.03$ & $+0.75$ & $-0.63$ & $-1.00/-1.00$ \\
\hline
\end{tabular}
\end{table*}

\section{Resolved polarimetric forward model and ray-tracing pipeline}
\label{app:rt}

\paragraph{Forward model.} For the tests of Sect.~\ref{sec:sign_ftomo} the
observer looks along $\hat{\mathbf{k}}=(0,\sin i,\cos i)$ in the frame in
which the jet axis is $\hat{\mathbf{z}}$. The sky axes are
$\hat{\mathbf{e}}_x=(1,0,0)$ and $\hat{\mathbf{e}}_v=(0,-\cos i,\sin i)$,
so that $i=90^\circ$ gives the side-on geometry of the main text. The field
and the electron density are resampled by trilinear interpolation (which
is differentiable) on an observer-aligned grid of $N\times N$ lines of sight
of $N$ cells each, long enough to cross the cube at any inclination. Along
each line of sight we accumulate
$I=\sum_s\varepsilon\,\Delta s$ and Eq.~(\ref{eq:ptomo}),
$P(\lambda^2)=-\sum_s\varepsilon\,e^{2i\chi_0}\,e^{2i\lambda^2\phi_s}\,\Delta s$,
where $e^{2i\chi_0}=(B_x+iB_v)^2/|\mathbf{B}_\perp|^2$ and $\phi_s$ is the
Faraday depth from the centre of cell $s$ to the observer. The emissivity is
$\varepsilon=|\mathbf{B}_\perp|^2$ for the synthetic tests. For ray-traced
images (below), $\varepsilon$ is a free positive field on the lattice. The
emission of a GRMHD flow follows the plasma, not $|\mathbf{B}|^2$, so the
field direction is then constrained by the polarization pattern and
$B_\parallel$ by the Faraday rotation, with $n_e\propto\varepsilon$. Three
global scales (total flux, fractional polarization, normalization of the
Faraday depth) are fitted with the field. The cost is the sum of the
normalized residuals of $I$ and of $P$ at each wavelength, plus the prior
and the divergence term of Sect.~\ref{sec:method}. Optimization uses Adam,
as in the main text. Relativistic aberration, Doppler boosting and light
bending are not included, which is appropriate for the
parsec-scale jet but not near the horizon.

\paragraph{Ray-traced images.} We render polarized images of the GRMHD
snapshots with \texttt{ipole} \citep{ipole2018}, which integrates the
polarized transfer equation with thermal synchrotron emission, absorption,
Faraday rotation and conversion along null geodesics of the Kerr metric.
The electron temperature follows the $R_{\rm high}$ prescription
($R_{\rm low}=1$, $R_{\rm high}=20$), and emission from regions with
$\sigma>10$ is excluded. ipole's default, $\sigma>1$, would also remove
the jet sheath that H-MOG reconstructs; the funnel interior, where the
density is set by numerical floors, remains excluded. The mass unit of each
snapshot is fixed so that the 230\,GHz flux is $0.5\,$Jy at $i=17^\circ$ ($M_{\rm BH}=6.2\times10^9\,M_\odot$,
$D=16.9\,$Mpc). Images are computed at four frequencies spanning the
1.3\,mm band (214, 230, 246 and 262\,GHz; MAD$_1$ only) and the 3\,mm band
(80, 86, 92 and 98\,GHz). Within each band the emitting region barely changes, so that
the differences between the images are dominated by Faraday rotation. The orientation of the images and the sign conventions of Stokes
$Q$, $U$ and of the Faraday rotation relative to our forward model were fixed
with controlled ipole runs on synthetic dumps: an off-axis emitting blob, and
blobs threaded by uniform fields of known direction. The scripts and
the test dumps are part of the released code.

\paragraph{Synthetic interferometric data.} The images are observed with the
\texttt{ngehtsim} package \citep{ngehtsim}, with the stations of the ngEHT
reference array \citep{ngeht2023}, typical weather for April, 2\,GHz bandwidth and thermal noise.
Two sets of maps are inverted. The first uses the true image convolved with
the clean beam fitted to the $(u,v)$ coverage, i.e.\ an ideal deconvolution at
the array resolution, which is a best case. The second reconstructs the
images by regularized maximum-likelihood imaging with \texttt{ehtim}
\citep{chael2018}: Stokes $I$ from visibility amplitudes and closure
quantities, then the linear polarization from the polarimetric visibilities,
on a $64\times64$ grid, after which the images are restored with the beam. All four frequencies are restored with a common circular beam,
the largest of the four fitted beams, and the model images are convolved with
the same beam, so that the differences between frequencies are not
produced by the frequency dependence of the resolution. The maps are then
inverted with H-MOG and compared with the true field of the snapshot.

\end{appendix}

\end{document}